\documentclass[aps,prd,twocolumn,superscriptaddress,floatfix,preprintnumbers]{revtex4} 

\usepackage{amsmath, amsthm, amssymb,bm,relsize}
\usepackage{amsmath,autobreak}
\usepackage{graphicx}  
\usepackage{amssymb}   
\usepackage{amsmath}
\usepackage{hyperref}
\usepackage{color}

\usepackage{caption}
\usepackage{booktabs,siunitx}

\allowdisplaybreaks
\usepackage{tabularx}
\newcolumntype{P}[1]{>{\centering\arraybackslash}p{#1}}
\usepackage{multirow, caption, booktabs}
\usepackage{slashed}
\usepackage{array}
\newcolumntype{P}[1]{>{\centering\arraybackslash}p{#1}}
\allowdisplaybreaks
\usepackage{endnotes}
\let\footnote=\endnote

\usepackage{multirow, caption, booktabs} 

\usepackage{graphicx}
\usepackage{psfrag}
\usepackage[caption=false]{subfig}
\definecolor{urlblue}{rgb}{0.2,0.4,0.7}
\definecolor{citegreen}{rgb}{0,0.6,0.2}
\definecolor{linkred}{rgb}{0.9,0.2,0.1}
\usepackage{hyperref}
\hypersetup{
colorlinks=true, citecolor=citegreen, linkcolor=blue,urlcolor=urlblue}

\usepackage{slashed}
\usepackage{array}
\newcolumntype{P}[1]{>{\centering\arraybackslash}p{#1}}

\allowdisplaybreaks

\def\zo{\overline{z}_1}
\def\zt{\overline{z}_2}

\def\w{\omega}

\usepackage{endnotes}
\let\footnote=\endnote

\begin{document}

\preprint{IMSc/2021/12/09} 
\preprint{BONN-TH-2022-02}

\title{Next-to-soft-virtual resummed rapidity distribution
for Drell-Yan process to $\rm{\textbf{NNLO}+\overline{\textbf{NNLL}}}$}

\author{A.H. Ajjath \footnote{ aabdulhameed@lpthe.jussieu.fr}}
\affiliation{The Institute of Mathematical Sciences, HBNI, Taramani,
 Chennai 600113, India}
 \affiliation{Laboratoire de Physique Th\'eorique et Hautes Energies (LPTHE), UMR 7589, Sorbonne Universit\'e et CNRS, 
4 place Jussieu, 75252 Paris Cedex 05, France}
\author{Pooja Mukherjee \footnote{pmukherj@uni-bonn.de}}
\affiliation{The Institute of Mathematical Sciences, HBNI, Taramani,
 Chennai 600113, India}
\affiliation{Bethe Center for Theoretical Physics, Universit\"at Bonn, D-53115, Germany}
\author{V. Ravindran\footnote{ravindra@imsc.res.in}}
\affiliation{The Institute of Mathematical Sciences, HBNI, Taramani,
 Chennai 600113, India}
\author{Aparna Sankar\footnote{aparnas@imsc.res.in}}
\affiliation{The Institute of Mathematical Sciences, HBNI, Taramani,
 Chennai 600113, India}
\author{Surabhi Tiwari\footnote{surabhit@imsc.res.in}}
\affiliation{The Institute of Mathematical Sciences, HBNI, Taramani,
 Chennai 600113, India}

\date{\today}

\begin{abstract}
We present the differential predictions for the rapidity distribution of a pair of leptons through the Drell-Yan (DY) process at the LHC taking into account the soft-virtual (SV) as well as next-to-soft virtual (NSV) resummation effects in QCD perturbation theory to next-to-next-to-leading-order plus next-to-next-to-leading- logarithmic ($\rm{{NNLO+\overline{NNLL}}}$) accuracy. We perform the resummation in two dimensional Mellin space using our recent formalism \cite{Ajjath:2020lwb} by limiting ourselves to contributions only from quark anti-quark ($q \bar q$) initiated channels. The resummed corrections to the fixed order results are computed through a matched formula using the minimal prescription procedure.  We find that the resummation at next-to-leading-logarithmic (next-to-next-to-leading-logarithmic) level brings about 3.98\% (1.24\%) corrections respectively to the NLO (NNLO) results at the central scale value of $q=M_Z$ for 13 TeV LHC. We also observe that the sensitivity to the renormalisation scale gets improved substantially by the inclusion of NSV resummed predictions at $\rm \overline{NNLL}$ accuracy. Further, the lack of quark gluon ($qg$) initiated 
contributions to NSV part in the  $\rm \overline{NNLL}$ resummed predictions leaves large
factorisation scale dependence indicating their importance at NSV level as we go to higher orders in perturbation theory. 

\end{abstract}

\maketitle

\section{Introduction}

The production of a pair of leptons, known as the
Drell-Yan (DY) production~\cite{Drell:1970wh}, is one of the well studied processes  
at TeV colliders such as Tevatron and the Large Hadron Collider (LHC).
This was possible due to wealth of precise theoretical predictions both in Standard Model (SM) and beyond SM 
taking into account corrections from various sources.  
Being a least contaminated process, DY production is used as
luminosity monitor~\cite{Khoze:2000db} at the LHC.
In addition, precise data on the rates allows one to fit the parton distribution functions of hadrons
\cite{Accardi:2016ndt,Scimemi:2017etj,Scimemi:2018xaf}.
Any deviation from the precise predictions can be used to set bounds on the parameters of models of new physics.

While the DY process at leading order (LO) is purely electroweak, the radiative corrections are dominated by
Quantum Chromodynamics (QCD) and it is an active area of interest for several decades, see~\cite{Altarelli:1978id,Altarelli:1979ub} for first next-to-leading order (NLO) results in perturbative QCD 
for DY process, and for invariant mass distribution of a pair of leptons at 
NNLO see ~\cite{Hamberg:1990np,Harlander:2002wh}.  For the same observable at  N$^3$LO level, 
the dominant soft-virtual (SV) contributions were obtained in ~\cite{Ahmed:2014cla,Li:2014bfa} prior
to complete result \cite{Duhr:2020seh} at N$^3$LO become available recently. 
Electroweak correction beyond LO can be found in ~\cite{Dittmaier:2001ay, Baur:2001ze}. 
In addition to invariant mass distribution, other differential distributions namely  rapidity, 
transverse momentum are known to N$^3$LO in QCD, see 
\cite{Ellis:1981hk,Anastasiou:2003yy,Anastasiou:2003ds,Catani:2009sm,Melnikov:2006kv,Ravindran:2006bu,Ravindran:2007sv,
Gavin:2012sy,Ahmed:2014uya,Chen:2021vtu}. 
For, mixed QCD and EW corrections see \cite{Li:2012wna} and  
for parton showers matched with NLO QCD results, see MC@NLO \cite{Frixione:2002ik}, POWHEG \cite{Frixione:2002ik,Frixione:2007vw} and
aMC@NLO \cite{Alwall:2014hca} frameworks.

The fixed order predictions are improved by resumming large threshold logarithms resulting from
soft gluons, see 
\cite{Sterman:1986aj,Catani:1989ne,Catani:1990rp,Moch:2005ky,Laenen:2005uz,Idilbi:2005ni,Ravindran:2005vv}.
For the transverse momentum distribution, at small $p_T$, the resulting large logarithms exponentiate in the
impact parameter space \cite{Collins:1984kg,Catani:2000vq}.  
In the soft-collinear effective theory (SCET)
one performs resummation in momentum space, see \cite{Idilbi:2005ky} for inclusive production and \cite{Becher:2010tm} for transverse momentum distribution.
In the Mellin space, resummation of large logarithms of the Feynman variable $x_F$ which describes the longitudinal 
momentum of the final state was achieved in \cite{Catani:1989ne} and it was found that there were two thresholds that could be resummed to all orders, also see~\cite{Westmark:2017uig} for a different scheme.  
In \cite{Mukherjee:2006uu} the resummation of rapidity of $W^\pm$ in the Mellin-Fourier (M-F) space was performed 
following a conjecture (see \cite{Laenen:1992ey})
and later on it was applied for Drell-Yan production in \cite{Bolzoni:2006ky,Bonvini:2010tp}.
A similar approach in SCET can be found in \cite{Becher:2006nr,Becher:2007ty}.

Following \cite{Catani:1989ne,Ravindran:2006bu,Ravindran:2007sv},
in \cite{Banerjee:2017cfc,Banerjee:2018vvb}, we studied the soft gluon resummation for the 
rapidity distributions of Higgs boson and also of a pair of leptons produced in hadron colliders.  
In the threshold limit, {\it i.e.}, when the scaling variables $z_1\to 1$ and $z_2\to 1$,  
the soft gluons contribute through delta functions and plus distributions 
in the partonic cross sections.
These contributions can be resummed to all orders both in $z_1,z_2$ space and in $N_1,N_2$ space.  
The resummed results known to desired logarithmic accuracy can be used
to predict certain highest logarithms in the fixed order, see ~\cite{Ravindran:2007sv,Ahmed:2014uya,Ahmed:2014era}.
The threshold limit denoted by ($z_1 \to 1$ , $z_2 \to 1$) corresponds to 
($N_1\to \infty$, $N_2\to \infty$) in the Mellin-Mellin (M-M) space, giving large logarithms of the 
form $\ln^n (N_i)$, where $n=1,\cdot \cdot \cdot$ and $i=1,2$ and the resummation in M-M space
resums terms of the form $\w=a_s \beta_0 \ln(N_1 N_2)$ through a process independent function $g(\w)$ and 
a process dependent but $N_i$ independent function $g_0$.   Here,
$a_s = g_s^2(\mu_R^2)/16 \pi^2$ with $g_s$ being the strong coupling constant and $\mu_R$, the renormalisation scale.  
The constant $\beta_0$ is the leading coefficient of the beta function in QCD.

Contrary to naive expectation, in certain inclusive \cite{Anastasiou:2015ema,Anastasiou:2015vya,Duhr:2019kwi,Duhr:2020seh} and differential \cite{Dulat:2018bfe} observables, 
one finds that the contributions from subleading threshold logarithms, called next to soft-virtual (NSV) terms 
contribute significantly at every order 
in perturbation theory.  They are found to be present in most of the partonic channels unlike the
leading logarithms.  Thanks to the availability of these fixed order results to unprecedented accuracy,  
there are enormous developments in the understanding of these subleading terms.  In particular,
questions related to their structure to all orders are still open,  see, \cite{Laenen:2008ux,Laenen:2010uz,Bonocore:2014wua,Bonocore:2015esa,Bonocore:2016awd,DelDuca:2017twk,Bahjat-Abbas:2019fqa,Soar:2009yh,Moch:2009hr,deFlorian:2014vta,Beneke:2018gvs,Beneke:2019mua,Beneke:2019oqx} for more details. Recently, in a series of articles \cite{Ajjath:2020ulr,Ajjath:2020sjk,Ajjath:2021bbm,Ajjath:2021lvg}, 
we studied a variety of inclusive observables to understand these subleading logarithms.  
Remarkably, we found that there exists a systematic way to sum them up
 to all orders in $z$ as well as in  the Mellin $N$ spaces, exactly the way one sums up leading threshold ones.  
This was achieved only for the diagonal channels.  
One finds that resummation of both SV and NSV terms can be achieved $N$ space.  
{\color{black}Later on, this was extended to study NSV terms  present in rapidity distributions \cite{Ajjath:2020lwb}
of a pair of leptons in DY and a Higgs boson in gluon fusion as well as in bottom quark annihilation. For  a generic case of $n$-colorless particles in the final state, see \cite{Ahmed:2020amh} for.  
Like the inclusive one, these subleading logarithms can be resummed to all order in 
multi-dimensional space (spanned by $z_l$ or $N_l$) along with the leading threshold logarithms }
In the present paper, we discuss the  phenomenological aspects  of resummed NSV terms for the production of a pair
of leptons at the LHC. In the subsequent sections, we briefly recapitulate the relevant theoretical results followed by a detailed study on the numerical impact of NSV contributions and finally we conclude our findings.
\section{Theoretical overview}
In the QCD improved parton model, the double differential distribution of a pair of leptons in DY process with respect to their  invariant mass $q^2$ and rapidity $y$ can be written as , 
\begin{eqnarray}\label{eq1}
{d^2 \sigma^q(\tau, q^2, y)\over dq^2 dy } &=&
\sigma^q_{\rm B}(x_1^0,x_2^0,q^2) 
\sum_{ab=q,\overline q, g}
\int_{x_1^0}^1 {dz_1 \over z_1}\int_{x_2^0}^1 {dz_2 \over z_2}~ 
\nonumber \\
&\!\!\hspace{-4.0cm}\times&
\hspace{-2.2cm}f_a\Big({x_1^0 \over z_1},\mu_F^2\Big)  f_b\Big({x_2^0\over z_2},\mu_F^2\Big)
\Delta^q_{d,ab} (z_1,z_2,q^2,\mu_F^2,\mu_R^2)\,,
\end{eqnarray} 
where $\sigma^q_{\rm B}(x_1^0,x_2^0,q^2) $ is the Born cross section, 
The dimensionless scaling variable $\tau$ is given by $\tau=q^2/S=x_1^0 x_2^0$ where $q$ being the momentum of
the pair of leptons and $S=(p_1 +p_2)^2$ in center of mass energy of incoming hadron with momenta $p_i$ respectively.
The rapidity $y$ of the lepton pair is given  by  $y=\frac{1}{2}\ln \Big(\frac{p_2.q}{p_1.q}\Big)=\frac{1}{2}\ln \left(\frac{x_1^0}{x_2^0}\right)$.  The parton distribution functions of incoming partons $a,b$ are non-perturbative and are denoted by  
$f_a\left({x_1^0 \over z_1},\mu_F^2\right)$ and   $f_b\left({x_2^0\over z_2},\mu_F^2\right)$ where ${x_1^0/z_1}$, ${x_2^0/z_2}$ are their momentum fractions respectively and are renormalised at the factorisation scale $\mu_F$.
$\Delta^q_{d,ab} (a_s,z_1,z_2,q^2,\mu_F^2)$ are the Drell-Yan coefficient functions (CFs) obtained from the partonic
subprocesses after mass factorisation at the scale $\mu_F$ and are 
calculable order by order in QCD perturbation theory in powers of $a_s$.
These CFs beyond leading order in perturbation theory contain distributions  
such as $\delta(1-z_i)$ and  
$\left[ \frac{\ln^{m-1}(1-z_i)}{1-z_i}\right]_+$ with $m\leq 2n$, $n$ being the order of
perturbation and regular functions of $z_i$.   Distributions show up only in the diagonal CFs, called 
SV terms and are denoted by $\Delta^{\text{SV}}_{d,q}$ while the regular part, also called the hard part is given by
$\Delta^{q,H}_{d,ab}$.  The leading terms in the hard part in the threshold expansion are nothing but the NSV terms.  Unlike the SV terms, the NSV terms get contributions from both diagonal as well as non-diagonal channels. Each term
belonging to NSV contribution contains a pair of either $\delta(1-z_i),i=1,2$ or ${\cal D}_l (z_i),i=1,2;l\ge 0$, where  ${\cal D}_l (z_i) = \left[ \frac{\ln^{l}(1-z_i)}{1-z_i}\right]_+$ and regular term $\log^k(1-z_j), j=1,2;k\ge 0$.
Following the work by Catani and Trentadue \cite{Catani:1989ne},
in \cite{Ravindran:2006bu,Ravindran:2007sv,Ahmed:2014uya}, the resummation  
of SV terms for the rapidity distribution 
to all orders was achieved in 
the scaling variables $z_i$ in $z_1,z_2$ space and 
later extended it to $N_1,N_2$ space in \cite{Banerjee:2017cfc,Banerjee:2018vvb} by performing 
two dimensional Mellin transformations 
in the large $N_1,N_2$ limit to obtain resummed result in the M-M space.  
In the Mellin $N_l$ space, when $N_l$ are large, the NSV terms take the form $\ln^k N_j/N_l$ with  $(j,l=1,2),(k=0,1\cdots$) up to $1/N_1^\alpha/N_2^\beta, \alpha,\beta \ge 1$.  In \cite{Ajjath:2020lwb}, restricting to diagonal channels, they were systematically resummed to all orders along with SV terms.

The diagonal CF, taking into account both SV and NSV, denoted by $\Delta^{\text{SV+NSV}}_{d,q}$ was shown to exponentiate in \cite{Ajjath:2020lwb} as 
\begin{align}\label{delta}
\Delta^{\rm SV+ \rm NSV}_{d,q} ={\cal C} \exp
\Big({\Psi^q_d(q^2,\mu_R^2,\mu_F^2,\zo,\zt,\epsilon)}\Big)\, \Big|_{\epsilon = 0} \,,
\end{align}
where the symbol ``$\mathcal{C}$" refers to convolution which acts on any exponential of a function $f(z)$ takes the
following expansion:
\begin{eqnarray}
        \mathcal{C}e^{f(z)} = \delta(1-z) + \frac{1}{1!}f(z) + \frac{1}{2!}\big(f\otimes f\big)(z) + \cdots
\end{eqnarray}
Here, we keep only  SV distributions, namely, $\delta(\bar{z}_l) \delta(\bar{z}_j)$,
$\delta(\bar{z}_l){\cal D}_i(z_j)$, ${\cal D}_i(z_l) {\cal D}_k(z_j)$, and NSV terms ${\cal D}_i(z_l)\ln^k(\bar{z}_j)$ and $\delta(\bar{z}_l)\ln^k(\bar{z}_j)$ with ($l,j=1,2)~(i,k=0,1,\cdots$) resulting from the convolutions.
In (\ref{delta}), $\epsilon$ is the complex valued parameter in the dimensional regularisation scheme. The function $\Psi^q_d$ in the above equation has the following integral representation in $z$-space


\begin{widetext}
\begin{align}
\label{eq:psiint}
\Psi^q_d =& {\delta(\overline z_1) \over 2} \Bigg(\!\!\displaystyle {\int_{\mu_F^2}^{q^2 \overline z_2}
\!\!{d \lambda^2 \over \lambda^2}}\! {\cal P}^q\left(a_s(\lambda^2),\zt\right) 
\!+\! {\cal Q}^q_d\left(a_s(q_2^2),\zt\right)
\!\!\Bigg)_+ 
+ {1 \over 4} \Bigg( {1 \over \overline z_1 }
\Bigg\{{\cal P}^q(a_s(q_{12}^2),\zt ) + {\color{black} 2 }L^q(a_s(q_{12}^2) ,\zt)
\nonumber \\
& + q^2{d \over dq^2} 
\left({\cal Q}^q_d(a_s(q_2^2 ),\zt) +  {\color{black} 2 }\varphi_{d,q}^f(a_s(q_2^2 ),\zt)\right)
\Bigg\}\Bigg)_+
+ {1 \over 2}
\delta(\overline z_1) \delta(\overline z_2)
\ln \Big(g^q_{d,0}(a_s(\mu_F^2))\Big)
+ \overline z_1 \leftrightarrow \overline z_2\,,
\end{align}
\end{widetext}

where $\bar{z}_l = 1- z_l$, $q_l^2 = q^2~(1-z_l) $, $q^2_{12}=q^2 \overline z_1 \overline z_2$ and the subscript $+$ indicates standard plus distribution. 

In (\ref{eq:psiint}), ${\cal P}^q (a_s, \overline z_l)= P^q(a_s,\overline z_l) - 2 B^q(a_s) \delta(\overline z_l)$ with $P^q(a_s,\overline z_l)$ being the splitting function in QCD which takes the following form,

\begin{align}\label{eq:Pq}
P^q(a_s,\overline z_l) = 2 \bigg({A^q (a_s) \over (\overline z_l)_+} +  B^q (a_s) \delta(\overline z_l) +  L^q(a_s, \overline z_l)\bigg)\,,
\end{align}
with $A^q$ and $B^q$ being the cusp and collinear anomalous dimensions, $L^q(a_s, \overline z_l) \equiv C^q(a_s) \ln(\overline z_l)  + D^q(a_s)$. 
The cusp ($A^q$), the collinear ($B^q$) anomalous dimensions and  the constants $C^q$ and $D^q$ 
to third order are available in \cite{Moch:2004pa,Vogt:2004mw,Blumlein:2021enk}. The anomalous dimensions $A^q, B^q, C^q $ and $D^q$ can also be found in the appendix of \cite{Ajjath:2021lvg}. In (\ref{eq:Pq}), we drop those terms which contribute to non-diagonal NSV and beyond NSV terms throughout. The function ${\cal Q}^q_d$ in (\ref{eq:psiint}) is given as
\begin{eqnarray}
{\cal Q}^q_d(a_s,\overline z_l) = {2 \over \overline z_l}  D_d^q(a_s) + 2 \varphi^f_{d,q} (a_s,\overline z_l)\,.
\end{eqnarray} 
In the above equation, the SV coefficient $ D_d^q$ are known to third order \cite{Banerjee:2017cfc} in QCD and the NSV coefficient $\varphi_{d,q}^{f}$ is parametrized in the following way,
\begin{align}
\label{eq:Phidf}
\varphi_{d,q}^f(a_s(\lambda^2),\overline z_l) &= \sum_{i=1}^\infty \sum_{k=0}^{\infty} \hat  a_s^i \left({\lambda^2 \over \mu^2}\right)^{i\frac{\epsilon}{2}}
S_\epsilon^i 
\varphi^{(i,k)}_{d,q}(\epsilon)\ln^k \overline z_l\,,
\nonumber\\
&= 
\sum_{i=1}^\infty \sum_{k=0}^i a_s^i(\lambda^2) \varphi^{q,(k)}_{d,i} \ln^k \overline z_l\,.
\end{align} 
The upper limit on the sum over $k$ is controlled by the dimensionally regularised Feynman integrals that contribute to order $a_s^i$.   
The constant $g^q_{d,0}$ in \eqref{eq:psiint} results from finite part of the virtual contributions and pure $\delta(\overline z_l)$ terms of real emission contributions. The NSV coefficients $\varphi^{q,(k)}_{d,i}$ in (\ref{eq:Phidf}) are known to third order \cite{Ajjath:2020lwb} and are listed below:

\begin{widetext}
\begin{align}
\label{eq:Phiq}
\varphi^{q,(0)}_{d,1} &= 
        2 C_F \,,
\nonumber \quad 
\varphi^{q,(1)}_{d,1} = 0\,,
\nonumber \quad
\varphi^{q,(0)}_{d,2} = C_F n_f   \bigg(  - \frac{268}{27}         + \frac{8}{3} \zeta_2 \bigg) + C_F C_A   \bigg(            \frac{1000}{27} - 28 \zeta_3 - \frac{56}{3} \zeta_2         \bigg) + C_F^2   \bigg(  - 16 \zeta_2 \bigg)\,,
\nonumber\\
\varphi^{q,(1)}_{d,2} &=
        C_F C_A   \bigg( 10 \bigg) + C_F^2   \bigg(  - 10 \bigg)\,,
\nonumber \quad
\varphi^{q,(2)}_{d,2} =
        C_F^2   \bigg(  - 4 \bigg)\,,
\nonumber\\
\varphi^{q,(0)}_{d,3} &=
        C_F n_f^2   \bigg( \frac{10856}{729} + \frac{32}{27} \zeta_3 - \frac{304}{27} \zeta_2 \bigg) + C_F C_A n_f   \bigg( - \frac{118984}{729} + \frac{196}{3} \zeta_3 + \frac{11816}{81} \zeta_2 - \frac{208}{15} 
         \zeta_2^2 \bigg) + C_F C_A^2   \bigg( \frac{587684}{729} \nonumber\\&
         + 192 \zeta_5 - \frac{21692}{27} \zeta_3 - \frac{40844}{81} \zeta_2
          + \frac{176}{3} \zeta_2 \zeta_3 + \frac{656}{15} \zeta_2^2 \bigg) + C_F^2 n_f   \bigg(  - \frac{1144}{9} + 96 \zeta_3 + \frac{1432}{27} \zeta_2 + \frac{32}{5} \zeta_2^2 \bigg) 
          \nonumber\\& + C_F^2 C_A   \bigg(  - \frac{5143}{27} + \frac{460}{9} \zeta_3 - \frac{5548}{27} \zeta_2 + \frac{1312}{15} 
         \zeta_2^2 \bigg) + C_F^3   \bigg( 23 + 48 \zeta_3 - \frac{32}{3} \zeta_2 - \frac{448}{15} \zeta_2^2 \bigg)\,,
\nonumber\\
\varphi^{q,(1)}_{d,3} &=
        C_F C_A n_f   \bigg(  - \frac{256}{9} - \frac{28}{9} \zeta_2 \bigg) + C_F C_A^2   \bigg( \frac{244}{9} + 24 \zeta_3 - \frac{8}{9} \zeta_2 \bigg) + C_F^2 n_f   \bigg( \frac{3952}{81} - \frac{64}{9} \zeta_2 \bigg) + C_F^2 C_A   \bigg(  - \frac{18436}{81} 
        \nonumber\\& + \frac{544}{3} \zeta_3 + \frac{436}{9} \zeta_2 \bigg) + C_F^3   \bigg( - \frac{64}{3} - 64 \zeta_3 + \frac{80}{3} \zeta_2 \bigg).
\nonumber\\
\varphi^{q,(2)}_{d,3} &=
        C_F C_A n_f   \bigg(  - \frac{10}{3} \bigg) + C_F C_A^2   \bigg( 34 - \frac{10}{3} \zeta_2 \bigg) + C_F^2 n_f   \bigg( \frac{40}{3} \bigg) + C_F^2 C_A   \bigg(  - 96 + \frac{52}{3} \zeta_2 \bigg) + C_F^3   \bigg( \frac{16}{3} \bigg)\,,
\nonumber\\
\varphi^{q,(3)}_{d,3} &=
        C_F^2 n_f   \bigg( \frac{32}{27} \bigg) + C_F^2 C_A   \bigg( - \frac{176}{27} \bigg)\,,
\end{align}
\end{widetext}
where the constants $C_A = N_c$ and $C_F = (N_c^2-1)/2 N_c$ are the Casimirs of $\rm{SU}(N_c)$ gauge group, $n_f$ is the number of active flavours and $\zeta_i$ are the Riemann zeta functions.

In \cite{Ajjath:2020lwb}, we systematically computed the analytic expression of resummed CFs in the M-M space after performing the double Mellin transformation on $\Delta_d^q$ in (\ref{delta}) as  
\begin{align} \label{eq:DeltaN}
\Delta_{d, N_1, N_2}^q = & \prod\limits_{i=1,2} \int_0^1 dz_i z_i^{N_i-1} \Delta_d^q(z_1,z_2) \\ \nonumber 
 = & \tilde g_{d,0}^q \exp(\Psi_{d,N_1, N_2}^q)\,,
\end{align}
where $\tilde g_{d,0}^q = \sum\limits_{i=0}^{\infty} a_s^i~ \tilde g_{d,0,i}^q$. Here, the resummed result for $\Psi_{d, N_1, N_2}^q$ takes the following form:
{\color{black}

\begin{align}
\label{eq:PsiN}
\Psi_{d, N_1, N_2}^q = &~~
  g_{d,1}^q(\omega)  \ln N_1
\nonumber\\&
+ \sum\limits_{i=0}^\infty a_s^i \bigg( \frac{1}{2}  g_{d,i+2}^q(\omega) + \frac{1}{N_1} \overline{g}_{d,i+1}^q(\omega) \bigg)
\nonumber\\&
 +\frac{1}{N_1} \left(h^q_{d,0}(\omega,N_1)  + 
\sum\limits_{i=1}^{\infty} a_s^i h^q_{d,i}(\omega,\omega_1,N_1)\right) 
\nonumber \\&
+ (N_1 \leftrightarrow N_2,\omega_1 \leftrightarrow \omega_2) \,,
\end{align}
}

where 

{\color{black}


\begin{align}
\label{hg}
h^q_{d,0}(\omega,N_l) &= h^q_{d,00}(\omega) + h^q_{d,01}(\omega) \ln N_l\,,
\nonumber \\ 
         h^q_{d,i}(\omega,\omega_l,N_l) &= \sum_{k=0}^{i-1} h^q_{d,ik}(\omega)~ \ln^k N_l 
         + \tilde{h}^q_{d,ii}(\omega,\omega_l)~ \ln^k N_l\,.
\end{align}
Here $\omega = a_s \beta_0 \ln N_1 N_2$,  $\omega_l = a_s \beta_0 \ln N_l$ for $l=1,2$ and $h^q_{d,01}(\omega) = 0$.

 The expressions in \eqref{eq:PsiN} and \eqref{hg} are slightly different from those in Eq.(12) and Eq.(13) of \cite{Ajjath:2020lwb}, due to the explicit $w_l$ dependence in the diagonal terms  $\tilde{h}^q_{d,ii}$, which needs an explanation. We notice that the diagonal terms $h^q_{d,ii}$ for $i \ge 2$ (or in general $h^c_{d,ii}$ for c=q,g,b which represent the Drell-Yan process, Higgs production in gluon fusion and bottom quark annihilation respectively) involve only the previous order information and hence can be included in the earlier order. Taking this fact into account, we redefine the diagonal terms at each order in the following way: 
\begin{align}
\tilde{h}^q_{d,11}(\omega,\omega_l) =& h^q_{d,11}(\omega) + \frac{\omega_l}{\beta_0} h^q_{d,22}
\nonumber \\
\tilde{h}^q_{d,ii}(\omega,\omega_l) = & \frac{\omega_l}{\beta_0} h^q_{d,i+1,i+1}\,, ~ \forall ~i \ge 2\,.
\end{align}
The results of $h^q_{d,ii}$ remain the same as in \cite{Ajjath:2020lwb}. Due to this rearrangement, the expression in \eqref{eq:PsiN} is better predictive as compared to the Eq.(12) of \cite{Ajjath:2020lwb}. This is clear when we compare Table \ref{tab:resNSV} in this paper with Table 1 of \cite{Ajjath:2020lwb}, where we depict the predictions for the NSV logarithms at higher order logarithmic accuracy using the lower order information. We emphasize that, both the definitions correctly predict the higher order resummed terms at every logarithmic accuracy and the only difference comes in how much lower order information are required for higher order resummed predictions. 
}

The SV resummation coefficients, which comprises of $\tilde g_{d,0}^q$ and $g_{d,i}^q$ are extensively discussed in \cite{Banerjee:2017cfc,Banerjee:2017ewt,Ahmed:2020caw}. The NSV coefficients  $\overline{g}_{d,i}^q$, $h^q_{d,ik}$ and $\tilde{h}^q_{d,ii}$ are listed in the appendices \ref{app:gbdN} and \ref{app:hdN}. Our next task is to include these resummed contributions consistently in the fixed order predictions to understand the phenomenological relevance of the NSV resummed results in the context of rapidity distribution for the lepton pair production in Drell-Yan process.
\section{Phenomenology}

In this section, we study the impact of resummed soft virtual plus next-to-soft virtual (SV+NSV) results for the rapidity distribution of a pair of leptons produced through $Z$ and $\gamma^*$ in the collision of two hadrons at the LHC with centre of mass energy 13 TeV. We take $n_f = 5$ flavors, the MMHT2014(68cl) PDF set \cite{Harland-Lang:2014zoa} 
and the corresponding $a_s(M_Z)$ through the Les Houches Accord PDF (LHAPDF) interface \cite{Buckley:2014ana} interface 
at each order in perturbation theory. For the fixed order rapidity distribution,
we use the publicly available code Vrap-0.9 \cite{Anastasiou:2003yy, Anastasiou:2003ds}. \textcolor{black}{The resummed contribution is obtained from $\Delta^{q}_{d,N_1,N_2}$ 
in (\ref{eq:DeltaN}) after performing Mellin inversions which
are done using an in-house FORTRAN based code.} The resummed results are matched to the fixed order result in order to avoid any double counting of threshold logarithms. 
The matched result is given below in (\ref{match}). Here $e_q$ is the charge of electron.
The numerical values for the various parameters are taken from the
Particle Data Group 2020 \cite{ParticleDataGroup:2020ssz} and are listed below :
\begin{align}\label{eq:parameters}
M_Z = 91.1876 ~ \text{GeV},  ~~ \Gamma_Z=2.4952~ \text{GeV} \,,\nonumber\\
 \text{s}^2_w=0.22343\,, ~~ \alpha=1/128 \,, ~~  c_w^2=1-s_w^2\,, \nonumber\\ g_e^V=-1/4+s_w^2\,\, ~~
g_u^V=1/4-2/3s_w^2\,\, \nonumber\\  g_d^V=-1/4+1/3s_w^2\,\, ~~ B^Z_{e} = 0.03363 \,.
\end{align}
\begin{widetext}
\begin{align}\label{match}
{d^2 \sigma^{q,\rm {N^nLO+\overline {\rm N^nLL}}} \over dq^2 dy } = &  
{d^2 \sigma^{q, \rm {N^nLO}} \over dq^2 dy } +
\, {\sigma^{q}_B } \sum_{ab\in\{q,\bar{q}\}} \int_{c_{1} - i\infty}^{c_1 + i\infty} \frac{d N_{1}}{2\pi i}
 \int_{c_{2} - i\infty}^{c_2 + i\infty} \frac{d N_{2}}{2\pi i} 
\left({\tau}\right)^{-N_{1}-N_{2}} 
\delta_{a b}f_{a,N_1}(\mu_F^2) f_{b,N_2}(\mu_F^2) \\ \nonumber &
\times \bigg( \Delta_{d,N_1,N_2}^q \bigg|_{\overline {\rm {N^nLL}}} - {\Delta^q_{d,N_1,N_2}}\bigg|_{tr ~\rm {N^nLO}} \bigg) \,,
\end{align}
where $\sigma^q_B$ is given by
\begin{eqnarray}
\label{matcha}
\sigma^{q}_B={4\pi \alpha^2 \over 3 q^4 N} \Bigg[e_q^2
-{2 q^2 (q^2-M_Z^2) e_q g_e^V g_q^V
\over
\left((q^2-M_Z^2)^2+M_Z^2 \Gamma_Z^2\right) c_w^2 s_w^2}
+{3 q^4 \Gamma_Z B^Z_e  \over  16 \alpha M_Z \left((q^2-M_Z^2)^2+M_Z^2 \Gamma_Z^2\right)
c_w^2 s_w^2 }\left(1+\Big(1-\frac{8}{3}s_w^2 \Big)^2 \right)\Bigg]\,.
\end{eqnarray}
\end{widetext}
\begin{table*}
\centering
\begin{small}
{\renewcommand{\arraystretch}{2.5}
\begin{tabular}{|P{1.6cm}||P{.05cm}P{2.4cm}P{2.9cm}P{2.9cm}P{0.8cm}P{3.5cm}|| P{1.7cm}|}
 \hline
 \hline
 \multicolumn{1}{|c||}{GIVEN} & & \multicolumn{4}{c}{PREDICTIONS - SV Logarithms} & & {Logarithmic Accuracy} \\
 \cline{1-1}\cline{2-7}\cline{1-1}
  Resummed exponents & &$\Delta^{q,(2)}_{d,N_1,N_2}$ & $\Delta^{q,(3)}_{d,N_1,N_2}$&$\Delta^{q,(4)}_{d,N_1,N_2}$& $\cdots$ &$\Delta^{q,(n)}_{d,N_1,N_2}$ &\\
 \hline
 	$\tilde  g^q_{d,0,0},g^q_{d,1}$ &  &  $\Big \{L_1^i L_2^j\Big \}{\Big|}_{i+j=4 }$ & $\Big \{L_1^i L_2^j \Big \}{\Big|}_{i+j=6}$ & $\Big \{L_1^i L_2^j \Big \}{\Big |}_{i+j=8}$ & $\cdots$&$\Big \{L_1^i L_2^j \Big \}{\Big |}_{i+j=2n}$&  ${\rm LL}$ \\
 $\tilde  g^q_{d,0,1},g^q_{d,2}$ &  &  &$\Big \{L_1^i L_2^j \Big \}{\Big |}_{ i+j=5,4}$ &$\Big \{L_1^i L_2^j \Big \}{\Big |}_{ i+j=7,6}$& $\cdots$ &$\Big \{L_1^i L_2^j \Big \}{\Big |}_{ i+j=2n-1,2n-2}$&  ${\rm NLL}$ \\ 
$\tilde  g^q_{d,0,2},g^q_{d,3}$ \newline  &  & & & $\Big \{L_1^i L_2^j\Big \}{\Big |}_{ i+j=5,4}$ & $\cdots$ & $\Big \{L_1^i L_2^j\Big \}{\Big |}_{ i+j=2n-3,2n-4}$  &  ${\rm NNLL}$ \\
 \hline 
 \hline
\end{tabular}}
\caption{\label{tab:resSV} The set of resummed exponents \Big\{$\tilde  g^q_{d,0,n},g^q_{d,n}$\Big\} which is required to predict the tower of SV logarithms in $\Delta_{d,N_1,N_2}^{q,(n)}$ at a given logarithmic accuracy in the Mellin $N$-space. Here,  $i,j \geq 0$ and $L^i_l = \ln^i N_l$ with $l=1,2$.}
\end{small}
\end{table*}
\begin{table*}
\centering
\begin{small}
{\renewcommand{\arraystretch}{2.5}
\begin{tabular}{|P{1.6cm}||P{.02cm}P{2.9cm}P{2.9cm}P{2.9cm}P{0.7cm}P{3.5cm}|| P{1.7cm}|}
 \hline
 \hline
 \multicolumn{1}{|c||}{GIVEN} & & \multicolumn{4}{c}{PREDICTIONS - NSV Logarithms} & & {Logarithmic Accuracy} \\
 \cline{1-1}\cline{2-7}\cline{1-1}
  Resummed exponents & &$\Delta^{q,(2)}_{d,N_1,N_2}$ & $\Delta^{q,(3)}_{d,N_1,N_2}$&$\Delta^{q,(4)}_{d,N_1,N_2}$& $\cdots$ &$\Delta^{q,(n)}_{d,N_1,N_2}$ &\\
 \hline
 		\multirow{2}{4em}{$\tilde g^q_{d,0,0},g^q_{d,1}, \newline  \overline g^q_{d,1},h^q_{d,0}$} \newline  &  &  $\Big \{L^{i,j}_{N_{1},2}, L^{i,j}_{N_{2},1} \Big \}{\Big|}_ {i+j=3 }$ & $\Big \{L^{i,j}_{N_{1},2}, L^{i,j}_{N_{2},1} \Big \}{\Big|}_ {i+j=5 }$  &$\Big \{L^{i,j}_{N_{1},2}, L^{i,j}_{N_{2},1} \Big \}{\Big|}_ {i+j=7 }$ &$\cdots$ &$\Big \{L^{i,j}_{N_{1},2}, L^{i,j}_{N_{2},1} \Big \}{\Big|}_ {i+j=2n-1}$&  $\overline{{\rm LL}}$\\
 \multirow{2}{4em}{$\tilde g^q_{d,0,1},g^q_{d,2},\newline \overline g^q_{d,2}, h^q_{d,1}$} \newline   &  &  &$\Big\{L^{i,j}_{N_{1},2}, L^{i,j}_{N_{2},1}\Big\}{\Big|}_ {i+j=4}$ &$\Big\{L^{i,j}_{N_{1},2}, L^{i,j}_{N_{2},1}\Big\}{\Big|}_ {i+j=6}$& $\cdots$ &$\Big \{L^{i,j}_{N_{1},2}, L^{i,j}_{N_{2},1} \Big \}{\Big|}_ {i+j=2n-2}$&  $\overline{{\rm NLL}}$ \\ 

\multirow{2}{4em}{$\tilde g^q_{d,0,2},g^q_{d,3},\newline \overline g^q_{d,3}, h^q_{d,2}$} \newline \newline   &  & & & $\Big \{L^{i,j}_{N_{1},2}, L^{i,j}_{N_{2},1}\Big \}{\Big|}_ {i+j=5}$ & $\cdots$ & $\Big \{L^{i,j}_{N_{1},2}, L^{i,j}_{N_{2},1} \Big \}{\Big|}_ {i+j=2n-3}$  &  $\overline{{\rm NNLL}}$ \\
 \hline 
 \hline
\end{tabular}}
\caption{\label{tab:resNSV} The set of resummed exponents \Big\{$\tilde g^q_{d,0,n},g^q_{d,n},\overline{g}^q_{d,n}, h^q_{d,n}$\Big\} which is required to predict the tower of NSV logarithms in $\Delta_{d,N_1,N_2}^{q,(n)}$ at a given logarithmic accuracy in the Mellin $N$-space. Here, $i,j \geq 0$, $L^{i,j}_{N_{1},2} = {\ln^i N_1 \over N_1} \ln^j N_2$ and $L^{i,j}_{N_{2},1} = {\ln^i N_2 \over N_2} \ln^j N_1$ .}
\end{small}
\end{table*}
The first term in (\ref{match}), 
$\left({d^2 \sigma^{q,\rm{N^nLO}} / dq^2 dy }\right)$, 
corresponds to contributions 
resulting from fixed order results up to $\rm{N^nLO}$.  
The second term on the other hand contains only SV and NSV logarithms but to all orders in perturbation theory. 
The subscript $``tr"$ in $\Delta^q_{d,N_1,N_2}$ indicates that it is 
truncated at the same order as the fixed order after expanding in powers of $a_s$.
Hence, at a given order $a_s^n$, the non-zero 
contribution from the second term starts at order $a_s^{n+1}$ and includes SV and NSV terms 
from higher orders. The Mellin space PDF ($f_{i,N}$) can be evolved using QCD-PEGASUS \cite{Vogt:2004ns}. 
Alternatively,  we use the technique described in \cite{Catani:1989ne,Catani:2003zt} to directly
deal with PDFs in the $z$ space. The contour $c_i$ in the Mellin inversion in (\ref{match})  can be chosen according to {\it Minimal prescription} \cite{Catani:1996yz} procedure. \textcolor{black}{Note that the same {\it Minimal prescription} procedure for the SV terms \cite{Catani:1996yz} will go through for the Mellin inversion of NSV terms as well. This is because the inclusion of NSV logarithms does not introduce any new Landau pole. Further, the Landau pole problem is directly related to the fact that the coupling $a_s$ enters into the non-perturbative region which results due to the integration of the argument of running coupling. And the position of the Landau pole is decided by the argument of $a_s$ in the integral representation (\ref{eq:psiint}) which is used to resum the large SV and NSV logarithms. Since the inclusion of NSV terms does not alter the argument of $a_s$ in (\ref{eq:psiint}) from the SV case, the Landau pole will remain same as that of SV resummation. In addition, as already noted in \cite{Catani:1996yz}, there will not be any power corrections induced for the SV resummation and this observation continues to be valid for the NSV case as well due to the {\it Minimal prescription} formula in (\ref{match}).} 

Our numerical results for the fixed order are based on NNLO computation of the CFs in which parton distribution functions are taken up to NNLO accuracy. \textcolor{black}{Although the results of fixed order rapidity distribution at N$^3$LO are presented in \cite{Chen:2021vtu} for the Drell-Yan process, the corresponding numerical code is not publicly available, therefore we could not include N$^3$LO results in our analysis.} The resummed SV and NSV results are computed up to NNLL accuracy. \textcolor{black}{To go beyond the NNLL accuracy, we need the collinear anomalous dimensions $C^q$ and $D^q$ to fourth order as well as the four loop NSV coefficient $\varphi_{d,4}^{q,4}$ which are currently not available}. To distinguish between SV and SV+NSV resummation, all along the paper, we denote the former by N$^n$LL and the latter by $\overline{\rm  N^nLL}$ for the $n^{\rm th}$ level logarithmic accuracy.

In Tables \ref{tab:resSV} and \ref{tab:resNSV}, we list the resummed exponents which are required to predict the tower of SV and NSV logarithms respectively in  $\Delta_{d,N_1,N_2}^q$ at a given logarithmic accuracy. Let us first discuss the predictions for the SV logarithms. As it is shown in Table \ref{tab:resSV}, using the first set of resummed exponents $\{\tilde g^q_{d.0,0}, g^q_{d,1}\}$ which constitute to the SV-${\rm{LL}}$ resummation, we get to predict the leading SV logarithms of $\Delta_{d,N_1,N_2}^{q,(i)}$ to all orders in perturbation theory, i.e, $\Big \{\ln^l N_1 \ln^k N_2 \Big \}$ with ${l+k=2i}~ (l,k \geq 0)$ at the order $a_s^i$ for all $i>1$. Further, using the second set of resummed exponents $\{\tilde g^q_{d,0,1},g^q_{d,2}\}$ along with the first set, one can predict extra the next-to-leading SV logarithms i.e the towers 
$\{\ln^l N_1 \ln^k N_2\}$ with $l+k= 2i-1, 2i-2$ for $\Delta_{d,N_1,N_2}^{q,(i)}$ with $i>2$. These towers of logarithms belong to the SV-${\rm{NLL}}$ resummation. In general, using the $n$-th set $\{\tilde g^q_{d,0,n},g^q_{d,n+1}\}$ in addition to the previous sets, we can predict the highest $(2n+1)$ towers of SV logarithms in $N_l$ with $l=1,2$, which constitutes to the SV-${\rm{N^nLL}}$ resummation, at every order in $a_s^{i}$ for all $i>n+1$, where $n=0,1,2 \cdots$.

Next, we discuss the predictions for the NSV logarithms present in $\Delta^{q}_{d,N_1,N_2}$ at a given logarithmic accuracy. As it is presented in Table \ref{tab:resNSV}, using the first set of resummed exponents $\{\tilde g^q_{d,0,0}, g^q_{d,1}, \overline g^q_{d,1}$, $h^q_{d,0}\}$ which constitute to the $\overline{\rm{LL}}$ resummation, we can predict the highest NSV logarithms of $\Delta_{d,N_1,N_2}^{q,(i)}$ to all orders in perturbation theory, i.e, $\{{\ln^l N_1 \over N_1} \ln^k N_2, {\ln^l N_2 \over N_2} \ln^k N_1\}$ with $l+k=2i-1$ at the order $a_s^i$ for all $i>1$. Similarly using the second set of resummation exponents $\{\tilde g^q_{d,0,1},g^q_{d,2},\overline g^q_{d,2}$, $h^q_{d,1}\}$ along with the first set, one can predict the next-to-highest NSV logarithms to all orders, which includes the towers  $\{{\ln^l N_1 \over N_1} \ln^k N_2, {\ln^l N_2 \over N_2} \ln^k N_1\}$ with $l+k=2i-2$ for $\Delta_{d,N_1,N_2}^{q,(i)}$ with $i>2$. These towers of logarithms contribute to the $\overline{\rm{NLL}}$ resummation. In general, using the $n$-th set $\{\tilde g^q_{d,0,n},g^q_{d,n+1},\overline g^q_{d,n+1}$, $h^q_{d,n}\}$ in addition to the previous sets, we get to predict the highest $(n+1)$ towers of NSV logarithms in $N_l$ with $l=1,2$, which constitute to the $\overline{\rm{N^nLL}}$ resummation, at every order in $a_s^{i}$ for all $i>n+1$. 

Below we present the resummed result given in \eqref{delta} at various logarithmic accuracy and discuss the resulting predictions for the NSV logarithms in $(N_1,N_2)$-space as displayed in Table \ref{tab:resNSV} till $a_s^4$ (N$^4$LO). Note that we set $\mu_R^2=\mu_F^2=q^2$ in the expressions of the predictions throughout.
We begin with the resummed result in the $\overline{\rm{LL}}$ approximation given by
\begin{widetext}
\begin{align}
\label{LL1}
    \Delta_{d,N_1,N_2}^{q,\overline{\rm{LL}}} &= \tilde g^q_{d,0,0} \exp\bigg[\ln N_1 ~ g^q_{d,1}(\omega)+ \frac{1}{N_1}\bigg(\overline g^q_{d,1}(\omega) +  h^q_{d,0}(\omega,N_1)\bigg)\bigg] + (N_1 \leftrightarrow N_2)\,.
\end{align}
Now we expand the above expression up to $a_s^4$ (N$^4$LO) and compare the predictions for leading NSV logarithms against those from fixed-order results. Note that, as can be seen from Table \ref{tab:resNSV}, the $\overline{\rm{LL}}$ resummation which comprises of only one loop anomalous dimensions and SV+NSV coefficients from  fixed-order NLO results, predicts the leading logarithms $\{{\ln^l N_1 \over N_1} \ln^k N_2, {\ln^l N_2 \over N_2} \ln^k N_1\}{\Big|}_ {l+k=3 }$,  $\{{\ln^l N_1 \over N_1} \ln^k N_2, {\ln^l N_2 \over N_2} \ln^k N_1\}{\Big|}_ {l+k=5 }, \{{\ln^l N_1 \over N_1} \ln^k N_2, {\ln^l N_2 \over N_2} \ln^k N_1\}{\Big|}_ {l+k=7 }$ etc at $a_s^2$ (NNLO), $a_s^3$ (N$^3$LO), $a_s^4$ (N$^4$LO) and so on respectively as shown below. At $a_s^2$ (NNLO), we have
\begin{align}
   \Delta_{d,N_1,N_2}^{q,(2)}|_{\rm{NSV}-\overline{\rm{LL}}}   &=   
       L^3_{N_{1}}  \bigg\{
           4  C_F^2
          \bigg\}
       +   L^{2,1}_{N_{1},2}  \bigg\{
           12  C_F^2
          \bigg\}
       + L^{1,2}_{N_{1},2}  \bigg\{
           12  C_F^2
          \bigg\}
       + {L^3_2\over N_1 }   \bigg\{
           4  C_F^2
          \bigg\}+ (N_1 \leftrightarrow N_2)  \,.
    \end{align}
At $a_s^3$ (N$^3$LO), we find
\begin{align}
   \Delta_{d,N_1,N_2}^{q,(3)}|_{\rm{NSV}-\overline{\rm{LL}}}   &=   
          L^5_{N_{1}}  \bigg\{
           4 C_F^3
          \bigg\}
       + L^{4,1}_{N_{1},2}     \bigg\{
           20 C_F^3
          \bigg\}
           +  L^{3,2}_{N_{1},2}  \bigg\{
           40 C_F^3
          \bigg\}
       +  L^{2,3}_{N_{1},2}  \bigg\{
           40 C_F^3
          \bigg\}
       +L^{1,4}_{N_{1},2} \bigg\{
           20 C_F^3
          \bigg\}
       + {L^5_2 \over N_1}   \bigg\{
           4 C_F^3 
          \bigg\} \\ \nonumber &
        +  (N_1 \leftrightarrow N_2)  \,.
    \end{align}
The prediction at $a_s^4$ (N$^4$LO) is 
\begin{align}
   \Delta_{d,N_1,N_2}^{q,(4)}|_{\rm{NSV}-\overline{\rm{LL}}}   &=   
     L^7_{N_{1}}  \bigg\{
           {8 \over 3} C_F^4
          \bigg\}

       + L^{6,1}_{N_{1},2}     \bigg\{
           {56 \over 3} C_F^4
          \bigg\}
       +  L^{5,2}_{N_{1},2}   \bigg\{
           56 C_F^4
          \bigg\}
       +L^{4,3}_{N_{1},2}   \bigg\{
           {280 \over 3} C_F^4
          \bigg\}
        +L^{3,4}_{N_{1},2}   \bigg\{
           {280 \over 3} C_F^4
          \bigg\}  \\ \nonumber &
        +L^{2,5}_{N_{1},2}   \bigg\{
           56 C_F^4
          \bigg\}
           +L^{1,6}_{N_{1}}   \bigg\{
           {56 \over 3} C_F^4
          \bigg\}
        + {L^7_2 \over N_1}   \bigg\{
           {8 \over 3} C_F^4 
          \bigg\}+ (N_1 \leftrightarrow N_2)  \,, 
    \end{align}
      where $L^{i,j}_{N_{1},2} = {\ln^i N_1 \over N_1} \ln^j N_2$, $L^k_{N_l} = {\ln^k N_l \over N_l}$ and $L^k_l = \ln N_l$ with $l=1,2$. Our predictions for the leading NSV logarithms are compared against the fixed-order results in \cite{Hamberg:1990np,Harlander:2002wh,Lustermans:2019cau} up to third order.
      
      Let us now turn to the resummed result at $\overline{\rm{NLL}}$ accuracy which reads as 
    \begin{align}
\label{NLL}
    \Delta_{d,N_1,N_2}^{q,\overline{\rm{NLL}}} &= ( \tilde g^q_{d,0,0} + a_s ~ \tilde g^q_{d,0,1} )~\exp\bigg[\ln N_1 ~ g^q_{d,1}(\omega) 
    +  g^q_{d,2}(\omega)   
    + \frac{1}{N_1}\bigg(\overline g^q_{d,1}(\omega) +  a_s ~\overline g^q_{d,2}(\omega)
  +  h^q_{d,0}(\omega,N_1)  \\ \nonumber & + a_s ~  h^q_{d,1}(\omega, \omega_1,N_1) \bigg)\bigg]  + (N_1 \leftrightarrow N_2, \omega_1 \leftrightarrow \omega_2) \,.
    \end{align} 
  Note that at $\overline{\rm{NLL}}$ accuracy, we require anomalous dimensions up to two loops and second order SV+NSV coefficients obtained from NNLO results.   
    After expanding the above result up to $a_s^4$, we obtain the predictions for the next-to leading NSV logarithms $\{{\ln^l N_1 \over N_1} \ln^k N_2, {\ln^l N_2 \over N_2} \ln^k N_1\}{\Big|}_ {l+k=4}$ , $\{{\ln^l N_1 \over N_1} \ln^k N_2, {\ln^l N_2 \over N_2} \ln^k N_1\}{\Big|}_ {l+k=6}$ etc at $a_s^3$, $a_s^4$ and so on respectively and are given by  
 \begin{align}
  \Delta_{d,N_1,N_2}^{q,(3)}|_{\rm{NSV}-\overline{\rm{NLL}}}   &~=     \Delta_{d,N_1,N_2}^{q,(3)}|_{\rm{NSV}-\overline{\rm{LL}}} 
  + L^4_{N_1} \bigg\{  \frac{110}{9} C_A C_F^2 
    - \frac{20}{9} C_F^2 n_f 
    + \Big(24 + 40 ~ \gamma_E \Big) C_F^3 \bigg\}
    + L^{3,1}_{N_1,2} \bigg\{ {440 \over 9} C_A C_F^2 \\ \nonumber &
     - {80 \over 9} C_F^2 n_f + \Big(  80 + 160 \gamma_E \Big) C_F^3 \bigg\}
      + L^{2,2}_{N_1,2}  \bigg\{ {220 \over 3} C_A C_F^2 
     - {40 \over 3} C_F^2 n_f + \Big(  88 + 240 \gamma_E \Big) C_F^3 \bigg\}\\ \nonumber &
      + L^{1,3}_{N_1,2}  \bigg\{ {440 \over 9} C_A C_F^2 
     - {80 \over 9} C_F^2 n_f + \Big(  32 + 160 \gamma_E \Big) C_F^3 \bigg\}
      + {L^4_2 \over N_1} \bigg\{ {110 \over 9} C_A C_F^2 
     - {20 \over 9} C_F^2 n_f + 40 \gamma_E  C_F^3 \bigg\}\\ \nonumber &
    + (N_1 \leftrightarrow N_2) \,,
    \end{align}
    \begin{align}
  \Delta_{d,N_1,N_2}^{q,(4)}|_{\rm{NSV}-\overline{\rm{NLL}}}   &~=     \Delta_{d,N_1,N_2}^{q,(4)}|_{\rm{NSV}-\overline{\rm{LL}}} 
  + L^6_{N_1} \bigg\{  \frac{154}{9} C_A C_F^3 
    - \frac{28}{9} C_F^3 n_f 
    + \Big(24 + {112 \over 3} ~ \gamma_E \Big) C_F^4 \bigg\}
    + L^{5,1}_{N_1,2}  \bigg\{ {308 \over 3} C_A C_F^3 \\ \nonumber &
     - {56 \over 3} C_F^3 n_f + \Big(  128 + 224 \gamma_E \Big) C_F^4 \bigg\}
      + L^{4,2}_{N_1,2} \Bigg\{ {770 \over 3} C_A C_F^3 
     - {140 \over 3} C_F^3 n_f + \Big(  272 + 560 \gamma_E \Big) C_F^4 \bigg\}\\ \nonumber &
      + L^{3,3}_{N_1,2}  \bigg\{ {3080 \over 9} C_A C_F^3 
     - {560 \over 9} C_F^3 n_f + \Big(  288 + {2240 \over 3} \gamma_E \Big) C_F^4 \bigg\}
      + L^{2,4}_{N_1,2}  \bigg\{ {770 \over 3} C_A C_F^3 
     - {140 \over 3} C_F^3 n_f\\ \nonumber &
     + \Big(  152 + 560 \gamma_E \Big) C_F^4 \bigg\}
      + L^{1,5}_{N_1,2}  \bigg\{ {308 \over 3} C_A C_F^3 
     - {56 \over 3} C_F^3 n_f + \Big(  32 + 224 \gamma_E \Big) C_F^4 \bigg\}\\ \nonumber &
      + {L^6_2 \over N_1} \bigg\{ {154 \over 9} C_A C_F^3 
     - {28 \over 9} C_F^3 n_f + {112 \over 3} \gamma_E  C_F^4 \bigg\}
    + (N_1 \leftrightarrow N_2) \,,
    \end{align}
    where $\gamma_E$ is the Euler-Mascheroni constant.
     The above predictions for the next-to leading NSV logarithms are in agreement with results given in \cite{Hamberg:1990np,Harlander:2002wh,Lustermans:2019cau} up to third order. \textcolor{black}{Furthermore, we have compared our full third order results for $\Delta_{d}^{q}(z_1,z_2)$ in $z$ space (see \cite{Ajjath:2020lwb} for the third order results in $z$ space) with the results obtained using the generalized threshold factorization approach presented in \cite{Lustermans:2019cau}. From that comparison, we have found that our third order
prediction is in complete agreement with results given in \cite{Lustermans:2019cau} for terms of the type ${\cal D}_i(z_l) \ln^j(\overline z_m)$, $i,j\ge 0 , l,m=1,2$ in the $z$ space. However, we could not compare the remaining $\delta(\overline z_l) \ln^j(\overline z_m)$ terms in our result because they were not available in \cite{Lustermans:2019cau}.  } 
 

Finally, using the  $\overline{\rm{NNLL}}$ resummation, which further embeds the three loop anomalous dimensions and third order SV+NSV coefficients obtained from N$^3$LO results, we predict the next-to-next-to leading logarithms $\{{\ln^l N_1 \over N_1} \ln^k N_2, {\ln^l N_2 \over N_2} \ln^k N_1\}{\Big|}_ {i+j=5}$  at $a_s^4$ (N$^4$LO), $\{{\ln^l N_1 \over N_1} \ln^k N_2, {\ln^l N_2 \over N_2} \ln^k N_1\}{\Big|}_ {i+j=7}$  at $a_s^5$ (N$^5$LO) and so on. The resummed expression at  $\overline{\rm{NNLL}}$ accuracy is given by
  \begin{align}
    \label{NNLL}
    \Delta_{d,N_1,N_2}^{q,\overline{\rm{NNLL}}} &= ( \tilde g^q_{d,0,0} + a_s ~ \tilde g^q_{d,0,1} +  a_s^2 ~ \tilde g^q_{d,0,2})
    \exp\bigg[\ln N_1 ~ g^q_{d,1}(\omega)
    +  g^q_{d,2}(\omega) + a_s ~ g^q_{d,3}(\omega) 
    + \frac{1}{N_1}\bigg(\overline g^q_{d,1}(\omega) +  a_s ~\overline g^q_{d,2}(\omega) 
     \\ \nonumber & + a_s^2 ~\overline g^q_{d,3}(\omega) 
    +  h^q_{d,0}(\omega,N_1) + a_s ~  h^q_{d,1}(\omega,\omega_1,N_1) 
    +  a_s^2 ~  h^q_{d,2}(\omega,\omega_1,N_1) \bigg)\bigg] + (N_1 \leftrightarrow N_2,\omega_1 \leftrightarrow \omega_2)  \,.
    \end{align}  
    The prediction for the next-to-next-to leading NSV logarithm at $a_s^4$ is provided below 
          \begin{align}
  \Delta_{d,N_1,N_2}^{q,(4)}|_{\rm{NSV}-{\overline{\rm{NNLL}}}}   &=    \Delta_{d,N_1,N_2}^{q,(4)}|_{\rm{NSV}-{\overline{\rm{NLL}}}} 
  + L^5_{N_1} \bigg\{ {968 \over 27} C_A^2 C_F^2 - {352 \over 27} n_f C_F^2 C_A + {32 \over 27} C_F^2 n_f^2 + \Big ({5392 \over 27} + {616 \over 3} \gamma_E  \\ \nonumber &
   - 24 \zeta_2 \Big) C_A C_F^3  - \Big ({1000 \over 27} + {112 \over 3}\gamma_E \Big ) C_F^3 n_f + \Big (-68 + 272 \gamma_E + 224 \gamma_E^2 + 64 \zeta_2 \Big ) C_F^4 \bigg\} \\ \nonumber &
 
  + L^{4,1}_{N_1,2}  \bigg\{ {4840 \over 27} C_A^2 C_F^2 
   -{1760 \over 27} n_f C_F^2 C_A 
   + {160 \over 27} C_F^2 n_f^2 
   + \Big({25916 \over 27} + {3080 \over 3}\gamma_E - 120 \zeta_2 \Big) C_A C_F^3 \\ \nonumber &
   - \Big({4712 \over 27} + {560 \over 3}\gamma_E \Big) C_F^3 n_f 
   + \Big(-360 + 1184 \gamma_E + 1120 \gamma_E^2 + 320 \zeta_2 \Big) C_F^4 \bigg\} 
   
   +  L^{3,2}_{N_1,2}  \bigg\{ {9680 \over 27} C_A^2 C_F^2 \\ \nonumber &
   -{3520 \over 27} n_f C_F^2 C_A 
   + {320 \over 27} C_F^2 n_f^2 
   + \Big({48952 \over 27} + {6160 \over 3}\gamma_E - 240 \zeta_2  \Big) C_A C_F^3 
   - \Big({8704 \over 27 } + {1120 \over 3}\gamma_E \Big) C_F^3 n_f \\ \nonumber &
   + \Big(-760 + 1952 \gamma_E + 2240 \gamma_E^2 + 640 \zeta_2 \Big) C_F^4  \bigg\}

   + L^{2,3}_{N_1,2} \bigg\{ {9680 \over 27} C_A^2 C_F^2 
   -{3520 \over 27} n_f C_F^2 C_A 
   + {320 \over 27} C_F^2 n_f^2   \\ \nonumber &
   + \Big(1664 + {6160 \over 3}\gamma_E - 240 \zeta_2  \Big) C_A C_F^3 
   - \Big(288 + {1120 \over 3}\gamma_E \Big) C_F^3 n_f 
   + \Big(-800 + 1472 \gamma_E + 2240 \gamma_E^2  \\ \nonumber &
   + 640 \zeta_2 \Big) C_F^4  \bigg\}
  
   + L^{1,4}_{N_1,2}  \bigg\{{4840 \over 27} C_A^2 C_F^2 
   -{1760 \over 27} n_f C_F^2 C_A 
   + {160 \over 27} C_F^2 n_f^2 
   + \Big({6568 \over 9} + {3080 \over 3}\gamma_E \\ \nonumber &
   -120 \zeta_2 \Big) C_A C_F^3 
   - \Big({1096 \over 9} + {560 \over 3}\gamma_E \Big) C_F^3 n_f 
   + \Big(-420 + 464 \gamma_E + 1120 \gamma_E^2 + 320 \zeta_2 \Big) C_F^4  \bigg\} 
  \\ \nonumber &
   + \frac{L^5_2}{N_1} \bigg\{ {968 \over 27} C_A^2 C_F^2 - {352 \over 27} n_f C_F^2 C_A + {32 \over 27} C_F^2 n_f^2 + \Big ({356 \over 3} + {616 \over 3} \gamma_E  
   - 24 \zeta_2 \Big) C_A C_F^3 \\ \nonumber &
   - \Big ({56 \over 3} + {112 \over 3}\gamma_E \Big ) C_F^3 n_f + \Big (-88 + 32 \gamma_E + 224 \gamma_E^2 + 64 \zeta_2 \Big ) C_F^4 \bigg\}\,.

    \end{align}
    \end{widetext}
The above predictions for the NSV logarithms in rapidity distribution $\Delta_{d,N_1,N_2}^{q,(i)}$ are found to reproduce the corresponding predictions in the inclusive cross section $\Delta_{N}^{q,(i)}$ computed in \cite{Ajjath:2021lvg} in the limit $N_1=N_2=N$ for $i \leq$4.

In \cite{Ajjath:2021lvg}, it has been shown, in the context of DY inclusive cross section up to N$^3$LO in the Mellin $N$ space that although the total contribution of NSV logarithms is smaller as compared to the SV counterpart, it is still numerically sizeable to not to be neglected.  Despite being formally subleading, the considerable contribution of the NSV terms is due to their large coefficients and this trend was also observed in the case of Higgs boson production through gloun fusion \cite{Anastasiou:2014lda,Ajjath:2021bbm}. Here, even in the case of rapidity distribution, we expect the same trend to be observed. Keeping this in mind, we ask the following questions that can shed light on the relevance of NSV terms in the context of rapidity distribution in di-lepton pair production in DY process at the LHC.
\begin{itemize}
\item
 In comparison to the fixed order corrections, how much is the effect of SV+NSV resummed results on the rapidity distribution?
\item
How sensitive is the SV+NSV resummed rapidity distribution to the choices of factorisation ($\mu_F$) and renormalisation  ($\mu_R$) scales?

\item 
How do the resummed NSV terms alter the predictions of SV resummed result?
\end{itemize}
In the following sections, we address the above questions in detail. Let us begin with analysing the impact of SV+NSV resummed results in comparison to the fixed order results, which is the topic of the next section. We present our results for the doubly-differential distribution with respect to invariant mass $q$ and rapidity $y$ by plotting it as a function of $y$ for fixed values of $q$. 
\subsection{Fixed order vs Resummed rapidity distribution}
\begin{table}
 \renewcommand{\arraystretch}{1.9}
\begin{tabular}{ |P{0.5cm}||P{1.3cm}|P{1.1cm}|P{1.5cm} |P{1.2cm}|P{1.9cm}|}
 \hline
  y &$\rm{K_{LO+\rm{\overline{LL}}}}$&$\rm{K_{NLO}}$&$\rm{K_{NLO+\rm{\overline{NLL}}}}$&$\rm{K_{NNLO}}$&$\rm{K_{NNLO+\rm{\overline{NNLL}}}}$\\
 \hline
\hline
 0 & 1.049 &  1.329 &1.382 &  1.369 &   1.386 \\
 0.8 &  1.05 &  1.319 & 1.372 &   1.358 &  1.374 \\
 1.6  & 1.05 &  1.291&  1.343 &  1.327 &  1.343 \\
 2.4  & 1.502 & 1.245 &  1.296 &1.279 &  1.295 \\
\hline
\end{tabular}
\caption{K-factor values of fixed order and resummed results at the central scale $\mu_R=\mu_F= M_Z$.} \label{tab:KmZ}
\end{table}
 In this section, we study the effects of SV+NSV resummation on the fixed order predictions for rapidity distribution of di-lepton pair production in DY process for 13 TeV LHC. Through (\ref{match}), we get the resummed predictions at $\rm{\overline{LL}}$, $\rm{\overline{NLL}}$ and $\rm{\overline{NNLL}}$ matched with the corresponding fixed order results. 
The numerical impact of higher order contributions can be quantified through the K-factors defined below, 
\begin{equation}\label{eq:Kfac}
 \mathrm{K} \left(q\right) = \dfrac{\dfrac{d^2\sigma}{dq^2 dy}\left(\mu_R=\mu_F=q\right)}{\dfrac{d^2\sigma^{\text{LO}}}{dq^2 dy}(\mu_R=\mu_F=q)} 
 \end{equation}
where we have set renormalisation ($\mu_R$) and factorisation ($\mu_F$) scales at $q$. 

\begin{figure}
\includegraphics[scale=0.47]{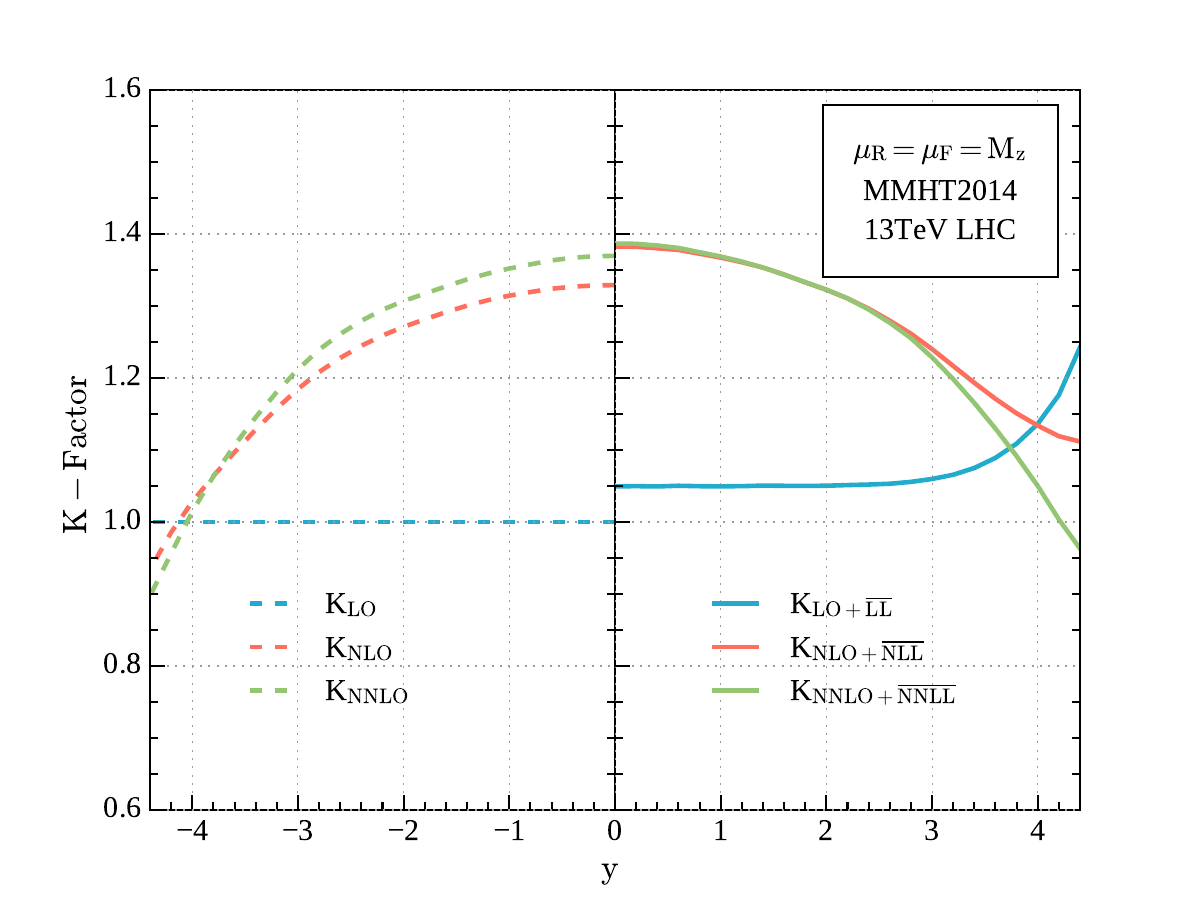}
\caption{The K-factor values for resummed results (right panel) in comparison to the fixed order ones (left panel) till NNLO+$\overline{\rm{NNLL}}$ level as a function of rapidity ($y$) at the central scale $\mu_R=\mu_F=M_Z$.}
\label{fig:KMz}
\end{figure}

Fig.(\ref{fig:KMz}) shows the K-factors at $\rm LO+{\overline{LL}}$, $\rm NLO+{\overline{NLL}}$ and $\rm NNLO+{\overline{NNLL}}$ in the right panel in comparison to the corresponding fixed order ones depicted in the left panel as a function of rapidity $y$ at the central scale $\mu_R=\mu_F=q$, where $q$ is fixed at $M_Z$.

\begin{table*}
 \renewcommand{\arraystretch}{2.7}
\begin{tabular}{ |P{0.6cm}||P{1.99cm}|P{1.99cm}|P{2.4cm}||P{2.3cm}|P{2.2cm}|P{2.2cm}|}
 \hline
  y &LO&NLO&NNLO&LO+$\rm{\overline{LL}}$&NLO+$\rm{\overline{NLL}}$&NNLO+$\rm{\overline{NNLL}}$\\
 \hline
\hline
 0 & 55.008$^{+ 14.99\%}_{-15.88\%}$ & 73.107$^{+ 2.951\%}_{-5.098\%}$ &  75.342$^{+0.6439\%}_{-0.9501\%}$&57.730$^{+ 15.36\%}_{-16.19\%}$& 76.049$^{+7.064\%}_{-7.502\%}$ & 76.283$^{+3.301\%}_{-2.178\%}$  \\
 0.8 & 54.674$^{+14.68\%}_{-15.59\%}$ & 72.137$^{+3.010\%}_{-5.083\%}$ & 74.237$^{+0.6864\%}_{-1.011\%}$&57.392$^{+ 15.05\%}_{-15.91\%}$& 75.044$^{+ 7.098\%}_{-7.482\%}$ & 75.159$^{+ 3.322\%}_{-2.233\%}$ \\
 1.6 & 53.293$^{+ 13.83\%}_{-14.80\%}$ &  68.825$^{+ 3.075\%}_{-4.988\%}$ &  70.735$^{+ 0.7423\%}_{-1.071\%}$& 55.972$^{+ 14.19\%}_{-15.11\%}$ &  71.607$^{+7.078\%}_{-7.370\%}$ &  71.600$^{+ 3.311\%}_{-2.278\%}$  \\
 2.4 &  50.327$^{+12.63\%}_{-13.71\%}$ &  62.642$^{+3.154\%}_{-4.866\%}$ & 64.392$^{+0.8200\%}_{-1.172\%}$ &52.944$^{+12.98\%}_{-14.02\%}$ &  65.238$^{+7.092\%}_{-7.251\%}$ &  65.182$^{+3.323\%}_{-2.370\%}$  \\
\hline
\end{tabular}
\caption{Values of resummed rapidity distribution at various orders in comparison to the fixed order results in pb/GeV at the central scale $\mu_R=\mu_F= M_Z$ for 13 TeV LHC.} \label{tab:FONSV7ptmZ}
\end{table*}

Below, in Table (\ref{tab:KmZ}), we show the K-factor values of both fixed order and resummed results for benchmark rapidity values at the central scale $\mu_R=\mu_F= M_Z$. We observe that there is an enhancement of $32.9\%$ and $36.9\%$ when we go from LO to NLO and NNLO respectively, at the central rapidity region. Furthermore, the fixed order values at LO, NLO and NNLO get incremented by the inclusion of $\rm SV+NSV$ resummed predictions at  $\rm{\overline{LL}}$, $\rm{\overline{NLL}}$ and $\rm{\overline{NNLL}}$ by $4.9\%$, 3.98\% and 1.24\% respectively, around the central rapidity region. This can be seen from the right panel of fig.(\ref{fig:KMz}), where the resummed curves are found to lie above their respective fixed order ones implying the enhancement resulting from the resummed corrections. It should be noted that the K-factor curves of the resummed results at $\rm NLO+{\overline{NLL}}$ and $\rm NNLO+{\overline{NNLL}}$ overlap for a wide range of rapidity values which was not observed for the case of fixed order predictions. This indicates that the perturbative convergence is improved among the resummed results, thereby leading to the reliability of perturbative predictions by the inclusion of resummed corrections. We also notice that the K-factor values are closer for NNLO and $\rm NNLO+{\overline{NNLL}}$ as compared to NLO and $\rm NLO+{\overline{NLL}}$ over the full rapidity region. This suggests that the resummed contributions to the fixed order rapidity distribution decrease as we go to higher orders in perturbation theory.       

From the above analysis of K-factors, we have observed that the resummed predictions not only bring in considerable enhancement in the fixed order results, but also improve the perturbative convergence till $\rm NNLO+{\overline{NNLL}}$ accuracy. However, both fixed order and resummed predictions suffer from the presence of unphysical scales, namely, the renormalisation $\mu_R$ and the factorisation $\mu_F$ scales.
Therefore, a careful study of perturbative uncertainties of these predictions is needed  by studying their sensitivity to the choices of $\mu_F$ and $\mu_R$ scales, which will be discussed in the following subsection.  

\subsubsection*{\textbf{7-point scale uncertainties
 of the resummed results}}\label{sec:7pt1}

\begin{figure*}
\includegraphics[scale=0.55]{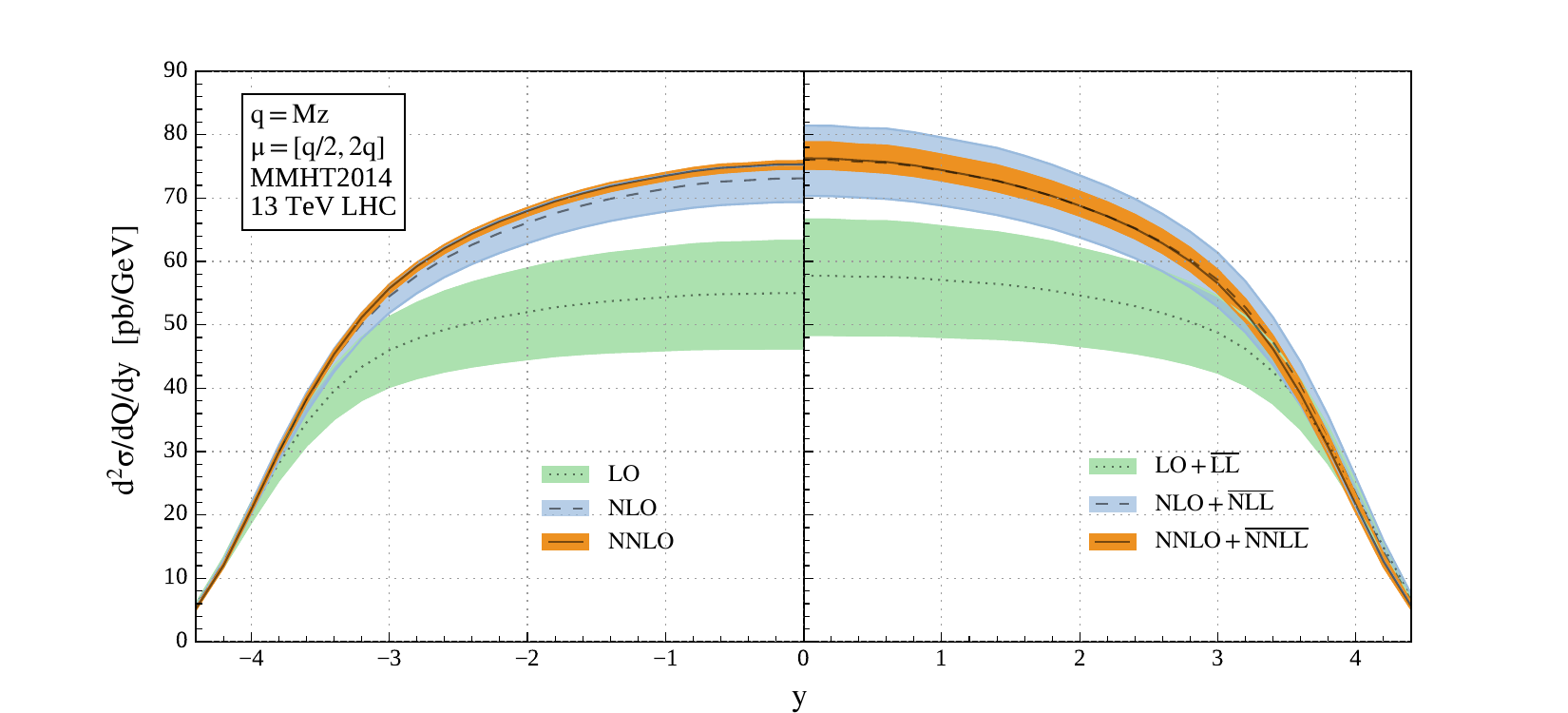}
\caption{7-point scale variation of the resummed result against fixed order around the central scale choice $(\mu_R,\mu_F) = ({M_Z,M_Z})$ for $13$ Tev LHC. The dotted, dashed and solid lines refer to the corresponding central scale values at each order.}
\label{fig:7ptMz}
\end{figure*}

The uncertainty associated with the choice of renormalisation $\mu_R$ and the factorisation $\mu_F$ scales due to the truncation of the perturbative series can be assessed using the standard canonical 7-point variation, where $\mu = \{ \mu_F, \mu_R \}$ is varied in the range $\frac{1}{2} \leq \frac{\mu}{q} \leq 2$, keeping the ratio $\mu_R/\mu_F$ not larger than 2 and smaller than 1/2.\\ 
In fig.(\ref{fig:7ptMz}), we compare the 7-point scale uncertainties of the $\rm SV+NSV$ resummed results (right panel) against fixed order ones (left panel) around the central scale choice $(\mu_R,\mu_F) = ({M_Z,M_Z})$ for $13$ Tev LHC at various perturbative orders. Here, we find that the central scale lines of resummed predictions are shifted up with respect to that of corresponding fixed order results. This indeed suggests that there is a systematic enhancement in the rapidity distribution when we add the resummed corrections to the fixed order results as shown in Table (\ref{tab:FONSV7ptmZ}). This was also observed from the analysis of K-factor values given earlier. However, we notice that the uncertainty bands of the resummed predictions are wider than that of the corresponding fixed order ones over the entire rapidity range at every order of perturbation. Nevertheless, the uncertainty band decreases as we go to higher logarithmic accuracy from $\rm LO+\rm \overline {LL}$ to $\rm NNLO+\rm \overline {NNLL}$. In addition, the error band of $\rm NNLO+\rm \overline {NNLL}$ is fully contained within the band of $\rm NLO+\rm \overline {NLL}$ over most of the rapidity region, unlike the fixed order case.\\  
In Table (\ref{tab:FONSV7ptmZ}), we present both fixed order and resummed predictions at various perturbative orders along with their asymmetric errors resulting from 7-point scale variation for benchmark rapidity values. Here, we notice an increment of 31.7\% while going from $\rm LO+\rm \overline {LL}$ to $\rm NLO+\rm \overline {NLL}$ accuracy, which further improves by 0.3\% at $\rm NNLO+\rm \overline {NNLL}$ for $y=0$. Besides this, the scale uncertainty gets reduced significantly while going from $\rm LO+\rm \overline {LL}$ to $\rm NNLO+\rm \overline {NNLL}$ over the full range of rapidity. For instance, the uncertainty ranges between $(-16.19 \%, +15.36 \%)$ for $\rm LO+\rm \overline {LL}$, $(-7.50 \%, +7.06 \%)$ for $\rm NLO+\rm \overline {NLL}$ and it is $(-2.18 \%, +3.30 \%)$ at $\rm NNLO+\rm \overline {NNLL}$ for central rapidity region. However, there is no improvement in the scale uncertainty of the resummed corrections when we compare it against the fixed order counterpart. 
This can be explained due to following possibilities:\\
\begin{itemize}
 \item   \textcolor{black}{The resummed logarithms are not dominant in this region to show their numerical relevance.} 
    
    \item  The lack of NSV resummed corrections from off-diagonal channels can give rise to large scale uncertainties in the resummed predictions.
\end{itemize}

\textcolor{black}{Recall that the resummation is inevitable to cure the perturbative series which suffers from certain large logarithmic terms at every order, in the kinematic threshold region, where the invariant mass $q$ approaches the hadronic centre of mass energy which is 13 TeV in our case.} Therefore, in order to see the impact of the resummed contributions, we need to extend our analysis to higher invariant mass region which is of the order of TeV. We will discuss the off-diagonal channel contribution later in detail towards the end of this section. Now, we move on to the analysis of 7-point scale uncertainty of the SV+NSV resummed predictions in comparison to the fixed order results for high invariant mass $q=$ 2 TeV.
\begin{table*}
 \renewcommand{\arraystretch}{2.7}
\begin{tabular}{ |P{0.6cm}||P{2.2cm}|P{2.2cm}|P{2.3cm}||P{2.2cm}|P{2.3cm}|P{2.4cm}|}
 \hline
  y &LO&NLO&NNLO&LO+$\rm{\overline{LL}}$&NLO+$\rm{\overline{NLL}}$&NNLO+$\rm{\overline{NNLL}}$ \\
 \hline
\hline
 0 &2.554$^{+7.627\%}_{-6.783\%}$ & 3.323$^{+ 2.836\%}_{-2.928\%}$ & 3.440$^{+0.597\%}_{-0.889\%}$ & 2.767$^{+ 8.026\%}_{-7.098\%}$ &3.460$^{+1.395\%}_{-1.253\%}$   & 3.470$^{+ 0.533\%}_{-0.312\%}$ \\

 0.4 & 2.385$^{+7.914\%}_{-7.011\%}$ & 3.105$^{+2.881\%}_{-2.998\%}$  & 3.233$^{+0.600\%}_{-0.919\%}$& 2.601$^{+8.360\%}_{-7.363\%}$  & 3.244$^{+ 1.408\%}_{-1.268\%}$  &   3.266 $^{+0.596\%}_{-0.397\%}$ \\
 
 0.8 & 1.762$^{+8.836\%}_{-7.720\%}$ & 2.295$^{+3.079\%}_{-3.290\%}$ & 2.409$^{+0.647\%}_{-1.033\%}$ & 1.962$^{+9.438\%}_{-8.185\%}$ & 2.426 $^{+1.563\%}_{-1.322\%}$ & 2.446$^{+0.7901\%}_{-0.653\%}$ \\
 
 1.2 &0.729$^{+10.72\%}_{-9.208\%}$ &  0.938$^{+3.655\%}_{-3.914\%}$ & 0.986$^{+0.788\%}_{-1.266\%}$ &0.851$^{+11.70\%}_{-9.918\%}$  &  1.019$^{+2.368\%}_{-1.458\%}$  &  1.013$^{+1.380\%}_{-1.206\%}$  \\
\hline
\end{tabular}
\caption{Values of resummed rapidity distribution at various orders in comparison to the fixed order results in $10^{-7}$ pb/GeV at the central scale $\mu_R=\mu_F= 2$ TeV for 13 TeV LHC.} \label{tab:FONSV7pt2T}
\end{table*}
 \begin{figure*}
\centering
\includegraphics[scale=0.55]{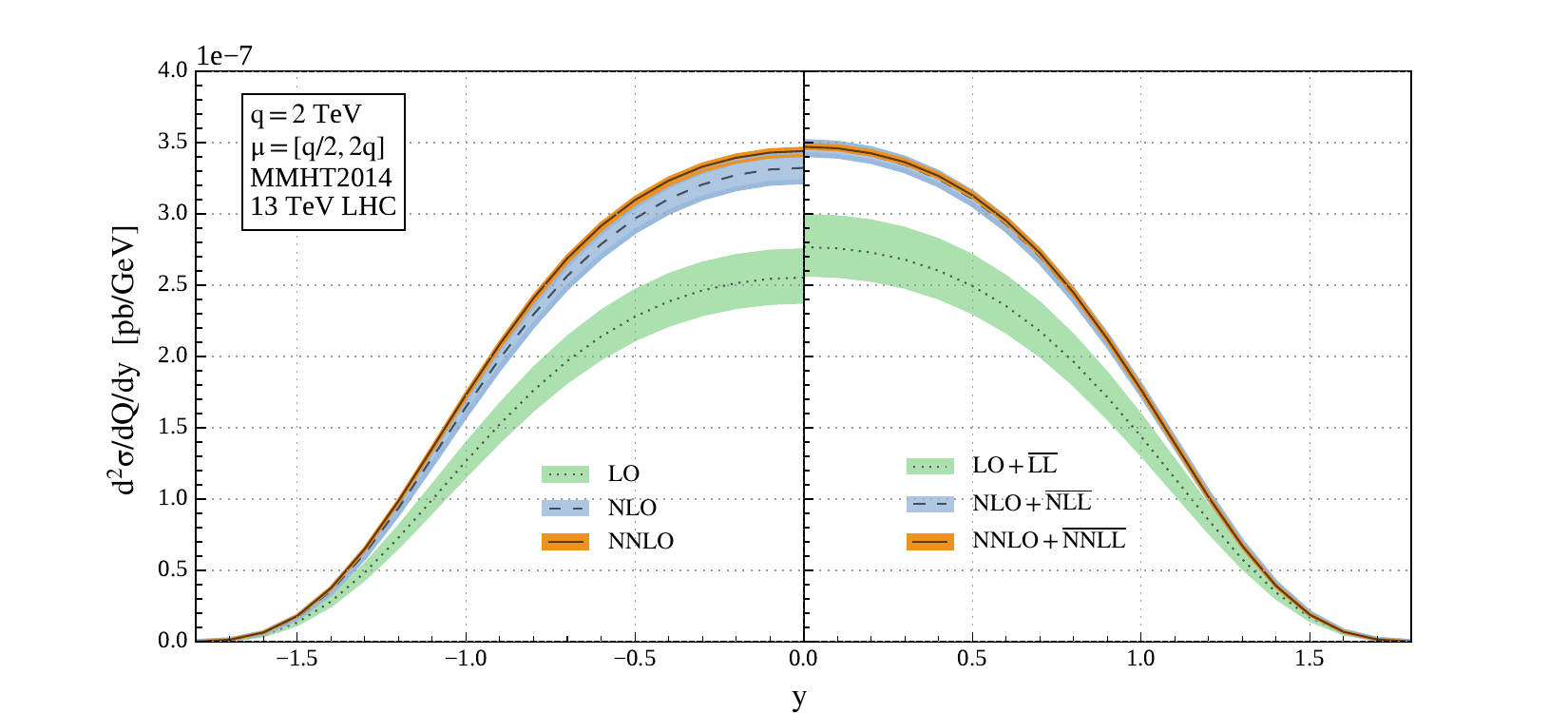}
\caption{7-point scale variation of the resummed result against fixed order around the central scale choice $(\mu_R,\mu_F) = (2,2)$ TeV for $13$ Tev LHC. The dotted, dashed and solid lines refer to the corresponding central scale values at each order.}
\label{fig:7pt2T}
\end{figure*}

From the earlier discussions on the 7-point scale uncertainty for $q=M_Z$, we found that the uncertainty bands of resummed predictions were wider than that of fixed order at every order of perturbation. Nevertheless, the width of uncertainty bands were found to decrease as we moved from $\rm LO+\rm \overline {LL}$ to $\rm NNLO+\rm \overline {NNLL}$ accuracy. In addition, we also observed appreciable amount of increment in the rapidity distribution by the inclusion of SV+NSV resummed effects. Now, here in fig.(\ref{fig:7pt2T}), we show the 7-point scale variation of the rapidity distribution for $q=2$ TeV. The fixed order results are depicted in the left panel up to NNLO accuracy and resummed predictions are given in the right panel up to $\rm NNLO+\rm \overline {NNLL}$ accuracy.
In general, we note that the width of uncertainty bands corresponding to both fixed order as well as resummed predictions are significantly
reduced as compared to the uncertainty bands for $q=M_Z$.
Interestingly, the $\rm NLO+\rm \overline {NLL}$ uncertainty band is better as compared to NLO fixed order band over the entire rapidity region. Also, the NNLO uncertainty gets improved by the inclusion of resummed  $\rm \overline {NNLL}$ corrections around the central rapidity region. This indicates the relevance of resummed contributions at this invariant mass region. This was not observed for the case of $q=M_Z$ where the resummed contributions were not prominent. 

In Table (\ref{tab:FONSV7pt2T}), we quote the central scale values of both fixed order and resummed rapidity distributions at $q=$2 TeV along with the 7-point scale uncertainties for benchmark rapidity values. Here, we observe that the percentage uncertainties of fixed order as well as resummed results get reduced substantially at each perturbative order when we compare them against the values given in Table(\ref{tab:FONSV7ptmZ}). For instance, the uncertainty at $\rm NNLO+\rm \overline {NNLL}$ is reduced from (-2.18\%,+3.3\%) to (-0.31\%,+0.53\%) as we go from $q=M_Z$ to $q=$ 2 TeV around the central rapidity region. In addition, the uncertainty at $\rm NNLO+\rm \overline {NNLL}$ is evidently small as compared to the uncertainty of (-0.89\%,+0.60\%) at NNLO for $y=0$. Similarly the uncertainty at NLO is (-2.93\%,+2.84\%) which comes down to (-1.25\%,+1.39\%) at  $\rm NLO+\rm \overline {NLL}$ for the same value of $y$. As for the case of $q=M_z$, there is a systematic reduction in the uncertainties while going from $\rm LO+\rm \overline {LL}$ to $\rm NNLO+\rm \overline {NNLL}$ over the entire rapidity region which can be seen from table (\ref{tab:FONSV7pt2T}). We also find that the resummed contribution at $\rm \overline {NNLL}$ brings in 0.86\% correction to NNLO whereas it was 1.24\% for the case of $q = M_Z$. This suggests that the correction resulting from resummation at $\rm \overline {NNLL}$ accuracy decreases as we go to higher $q$ values leading to better reliability of resummed results. \\
To summarise, we found that the uncertainties of the rapidity distribution decrease by the inclusion of the resummed corrections at $q=2$ TeV over the full rapidity region. Furthermore, the reliability of the perturbative results due to resummed corrections is improved at this invariant mass value. Thus, it can be inferred that the relevance of resummation effects become evidently visible while going from $q=M_z$ to $q=2$ TeV. To understand these observations in a better way, we now turn to study the effect of each scale individually on the SV+NSV resummed result.    

\subsubsection*{\textbf{Uncertainties of the resummed results \textit{with respect to}  $\mu_R$ and $\mu_F$ }}\label{sec:muRmuF1}

\begin{figure*}[!ht]
\includegraphics[scale=0.55]{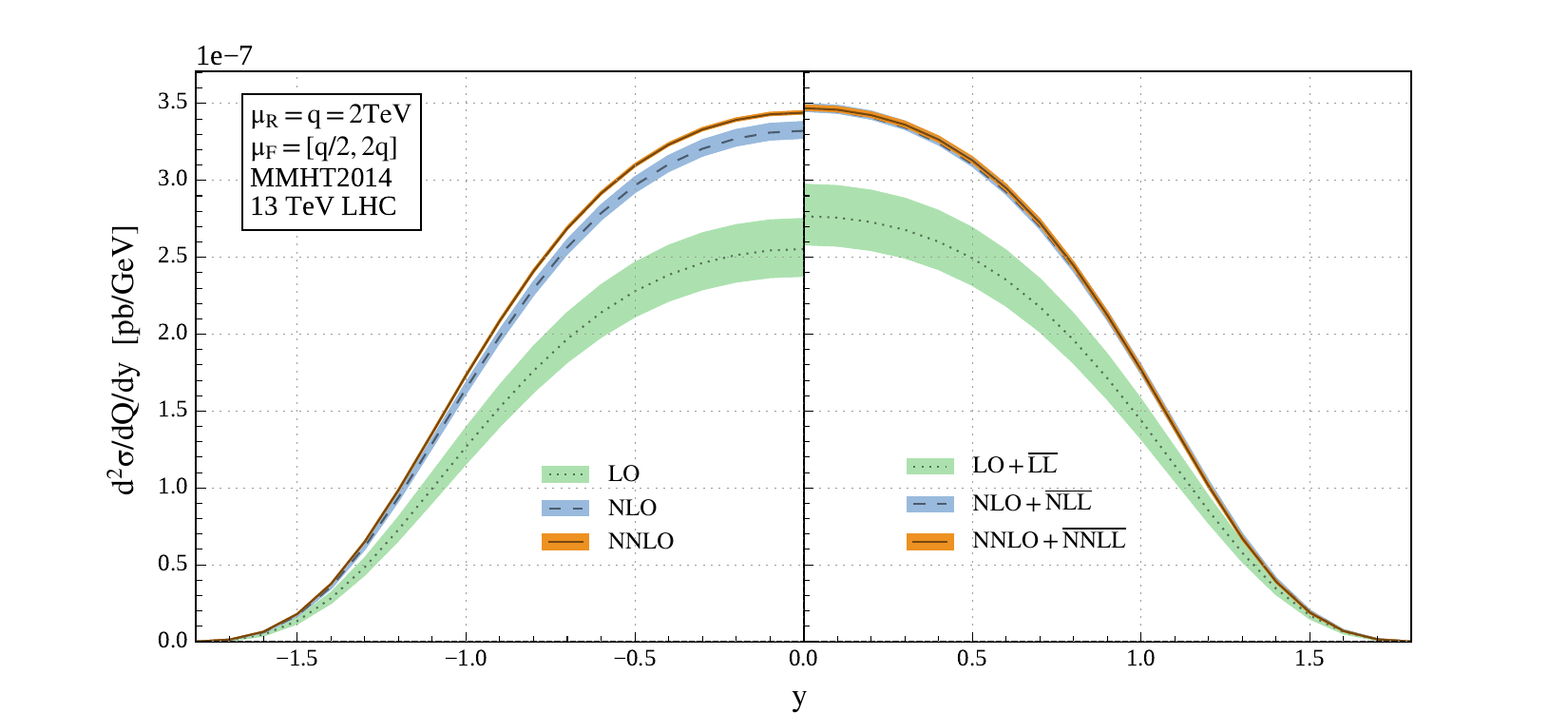}
\caption{$\mu_F$ scale variation of the resummed results against the fixed order with the scale $\mu_R$ held fixed at 2 TeV for 13 TeV LHC. The dotted, dashed and solid lines refer to the corresponding central scale values at each order.}
\label{fig:muF2T}
\end{figure*}

In the following, we examine the effect of $\mu_R$ and $\mu_F$ scales individually on the resummed result. We begin with plotting the dependence of the rapidity distribution on $\mu_F$ as a function of the rapidity $y$ while fixing the scale $\mu_R$ at the invariant mass $q$, for $q=2$ TeV as shown in fig.(\ref{fig:muF2T}). The bands are obtained by varying the scale $\mu_F$ by a factor of two up and down around the central scale $\mu_R=\mu_F=2$ TeV. Here, the resummed band depicted in the right panel at NNLO+$\overline{\rm NNLL}$ looks similar to that of the 7-point variation band shown in fig.(\ref{fig:7pt2T}) (right panel). This indicates that the contribution to the width of NNLO+$\overline{\rm NNLL}$ band in fig.(\ref{fig:7pt2T}) mainly comes from the uncertainties arising from variations in the $\mu_F$ scale. 
Note that the uncertainties at NNLO+$\overline{\rm NNLL}$ arising from $\mu_F$ and 7-point variation are identical and they lie between (-0.31\%, +0.53\%) for $y=0$. Moreover, the $\mu_F$ scale uncertainties decrease as we go to higher logarithmic accuracy in the resummed results.  

Now, we move on to compare the $\mu_F$ scale uncertainty of the resummed predictions with respect to the fixed order results. We observe that the $\mu_F$ scale uncertainty of NLO gets improved by the inclusion of $\overline{\rm NLL}$ resummed predictions whereas the NNLO band increases when the $\overline{\rm NNLL}$ corrections are added. Let us try to understand why the SV+NSV resummed result at NNLO+$\overline{\rm NNLL}$ is more sensitive to the $\mu_F$ scale variation as compared to fixed order NNLO result.
As mentioned earlier, we perform the resummation of SV distributions and NSV logarithms present in the diagonal partonic channel. Unlike the SV distributions that get contribution only from diagonal quark anti-quark ($q\bar{q}$) initiated channel, the NSV terms can originate from off-diagonal channels like quark gluon ($qg$), gluon gluon ($gg$) etc as well. Under the $\mu_F$ scale variation, these various partonic channels get mixed due to the DGLAP evolution of the PDFs. Hence, it becomes essential to keep all the contributing partonic channels at a particular perturbative order as there can be compensations among those channels, thereby reducing the scale uncertainty at that order. The fixed order results used for our numerical analysis contain all the partonic channels while the resummed contributions are only from $q\bar{q}$ initiated  channels. Thus, the scale dependence of fixed order result is expected to go down in comparison to the corresponding resummed prediction. 

However, as mentioned above, the inclusion of resummed corrections at $\rm \overline {NLL}$ accuracy improves the NLO error band. This suggests that the contribution of $qg$ channel is not prominent at NLO. We find that the one-loop correction from the $q\bar q$-channel is 23.6\% while the correction from the $qg$-channel is only -2.5\% of the NLO rapidity distribution at the central rapidity value. Therefore, there is an improvement in the $\mu_F$ scale uncertainty when we sum up the collinear logarithms resulting from the dominant $q\bar{q}$-channel at $\overline{\rm NLL}$.  
On the other hand, at NNLO level, the $a_s^2$ corrections from $q\bar q$ and $qg$ channels are $4.5\%$  and $-1.25\%$ respectively to the NNLO rapidity distribution. As a result, the magnitude of the NNLO result is determined by a significant cancellation between $q\bar{q}$ and $qg$ channels which was not the case for NLO.
Now, due to the unavailability of the $qg$ resummed collinear logarithms in our analysis, the aforementioned cancellation at NNLO+$\overline{\rm NNLL}$ is not balanced. Thus, the $\mu_F$ variation band of resummed prediction at NNLO+$\overline{\rm NNLL}$ in fig.( \ref{fig:muF2T}) displays that the $qg$ resummation is required to improve the results.  
\begin{figure*}
\includegraphics[scale=0.55]{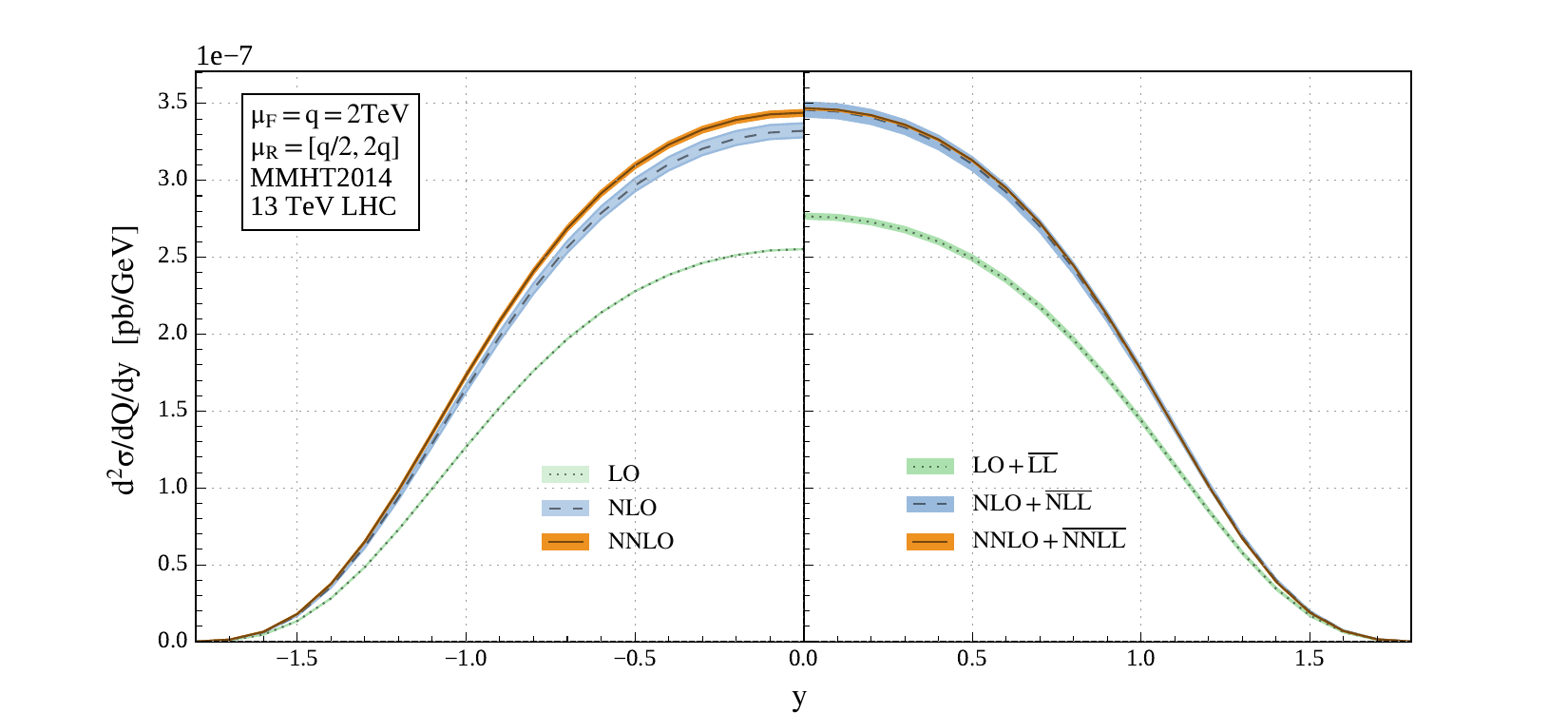}
\caption{$\mu_R$ scale variation of the resummed results against the fixed order with the scale $\mu_F$ held fixed at 2 TeV for 13 TeV LHC. The dotted, dashed and solid lines refer to the corresponding central scale values at each order.}
\label{fig:muR2T}
\end{figure*}

Next we try to understand the behaviour of resummed rapidity distribution in comparison to the fixed order counterpart under $\mu_R$ scale variation. Fig.(\ref{fig:muR2T}) shows the dependence of the rapidity distribution on $\mu_R$ keeping $\mu_F$ fixed at 2 TeV. The bands are obtained by varying the scale $\mu_R$ by a factor of two up and down around the central scale $\mu_R=\mu_F=2$ TeV. \textcolor{black}{Here, the LO rapidity distribution being independent of the scale $\mu_R$, does not have a band associated with it. On the other hand, there is a band when we add the resummed corrections at $\overline{\rm LL}$} accuracy to the LO rapidity distribution. This is because the resummed corrections at $\overline{\rm LL}$ capture the leading logarithmic contributions from all orders in perturbation theory, thereby giving rise to $\mu_R$ scale uncertainty. Moreover, the inclusion of resummed corrections at both $\overline{\rm NLL}$ and $\overline{\rm NNLL}$ improves the $\mu_R$ scale uncertainties of NLO and NNLO respectively. This is in contrast to the case of $\mu_F$ scale variation discussed earlier. Although the improvement is minuscule at NLO, it is substantial at NNLO due to NNLO+$\overline{\rm NNLL}$ which is indeed the highlight of this plot here as compared to $\mu_F$ scale variations shown in fig.(\ref{fig:muF2T}).
For instance, the $\mu_R$ scale uncertainty at NNLO is reduced from (-0.56\%, +0.5\%) to (-0.16\%, 0\%) for $y=0$ by the inclusion of $\overline{\rm NNLL}$. As we know, each partonic channel is invariant under $\mu_R$ scale variation when taken to all orders. Hence, there is an improvement when we include more higher order corrections within a channel which is $q\bar{q}$ in this case, by keeping the scale $\mu_F$ fixed. 

In conclusion, we observed that the uncertainties due to both $\mu_F$ and $\mu_R$ scales decrease as we go to higher logarithmic accuracy. As far as the $\mu_F$ scale variation is concerned, the resummation of collinear logarithms resulting from $qg$ channel also plays an important role. We notice that having the $qg$ resummed contribution is more significant at NNLO level than at NLO due to relatively larger contribution from $qg$ channel at NNLO. As a result, the 7-point scale uncertainty of the SV+NSV resummed predictions at NNLO+$\overline{\rm NNLL}$ is mostly driven by the $\mu_F$ scale variation. Note that the inclusion of SV+NSV resummed predictions reduces the the $\mu_R$ scale sensitivity remarkably at NNLO+$\overline{\rm NNLL}$ accuracy.           
So far we have discussed the effects of resummation on the fixed order results taking into account SV distributions and NSV logarithms together in the analysis. Now let us turn to understand which part of the SV+NSV resummation, i.e., whether it is the resummation of the distributions or of the NSV logarithms, plays the main role in any kind of improvement observed so far.
\subsection{SV vs SV+NSV Resummed results}
\begin{figure}
\includegraphics[scale=0.47]{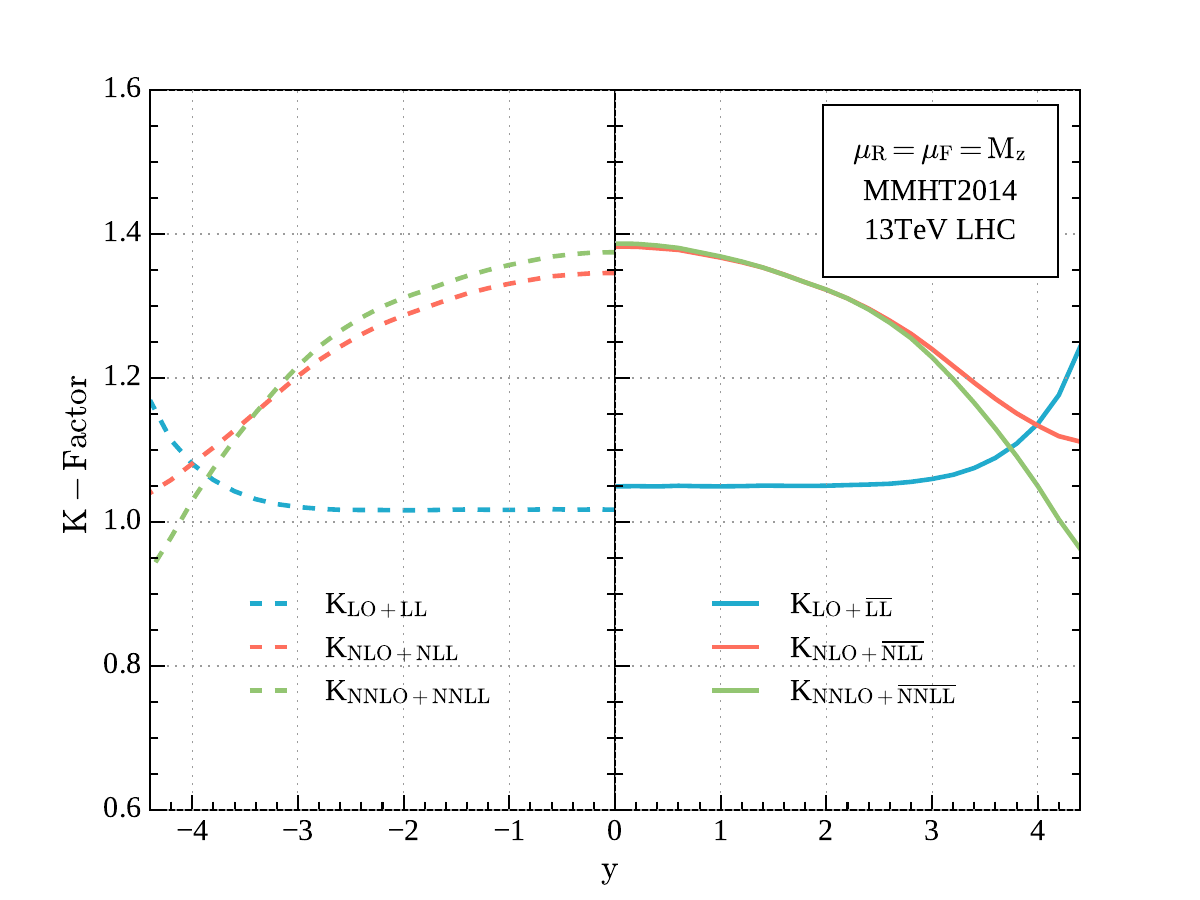}
\caption{The K-factor values for SV+NSV resummed results in comparison to the SV ones till NNLO+$\overline{\rm{NNLL}}$ level at the central scale $\mu_R=\mu_F= M_Z$.}
\label{fig:KsvnsvMz}
\end{figure}
\begin{table*}
 \renewcommand{\arraystretch}{1.7}
\begin{tabular}{ |P{0.4cm}||P{1.8cm}|P{1.8cm}|P{2.0cm} |P{1.9cm}|P{2.0cm}|P{2.0cm}|}
 \hline
  y &$\rm{K_{LO+\rm{{LL}}}}$&$\rm{K_{LO+\rm{\overline{LL}}}}$&$\rm{K_{NLO+\rm{{NLL}}}}$&$\rm{K_{NLO+\rm{\overline{NLL}}}}$ &$\rm{K_{NNLO+\rm{{NNLL}}}}$&$\rm{K_{NNLO+\rm{\overline{NNLL}}}}$  \\
 \hline
\hline
 0 & 1.017  & 1.049  & 1.345 &  1.382 &  1.374  & 1.386\\
 0.8 & 1.017  &  1.05 & 1.336 &  1.372 & 1.362  & 1.374\\
 1.6  & 1.017 &  1.05 & 1.307  & 1.343 & 1.332 &  1.343\\
 2.4  & 1.016 &  1.05 & 1.260 &  1.296 & 1.283 & 1.295\\

\hline
\end{tabular}
\caption{The K-factor values for SV+NSV resummed results in comparison to the SV results till NNLO+$\overline{\rm{NNLL}}$ level at the central scale $\mu_R=\mu_F= M_Z$.} \label{tab:KsvnsvmZ}
\end{table*}
\begin{figure*}
\includegraphics[scale=0.55]{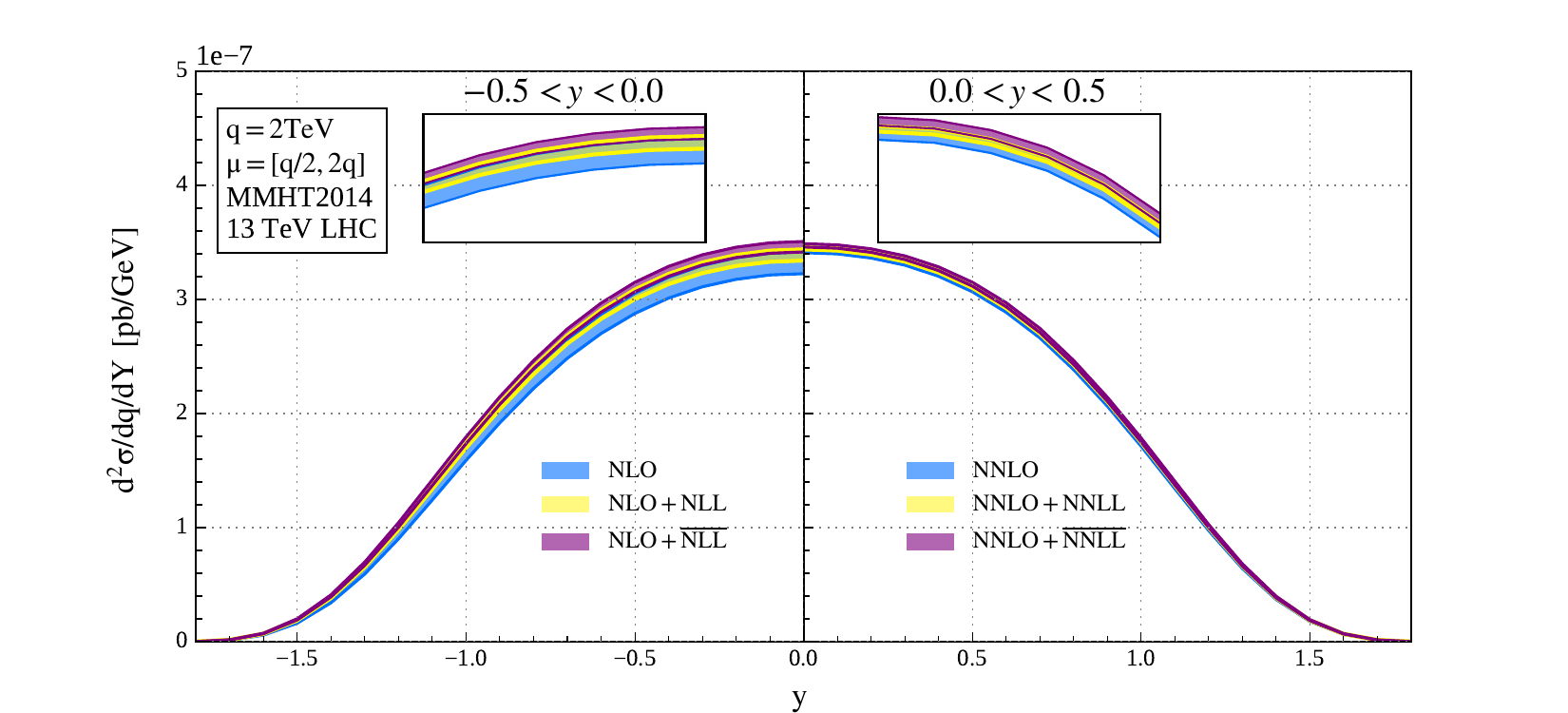}
\caption{Comparison of 7-point scale variation between SV and SV+NSV resummed results matched to NLO(left panel) and NNLO(right panel) for $q=2$ TeV.}
\label{fig:7pt2Tsvnsv}
\end{figure*}
In the previous section, we have studied the effects of SV+NSV resummation on the fixed order rapidity distribution in detail. We observed that there is a considerable amount of  enhancement in the rapidity distribution by the inclusion of SV+NSV resummed predictions and more importantly the $\mu_R$ scale uncertainty gets reduced substantially at NNLO+$\overline{\rm NNLL}$ accuracy. On the other hand, the $\mu_F$ scale uncertainty shows improvement at NLO+$\overline{\rm NLL}$ for higher values of $q$ but not at NNLO+$\overline{\rm NNLL}$. In the following, we perform an analysis on the inclusion of resummed NSV logarithms by comparing it with the SV resummed results.

We begin with the analysis of K-factor values for SV+NSV resummed results in comparison to the SV counterpart till NNLO+$\overline{\rm{NNLL}}$ level at the central scale $\mu_R=\mu_F=q$ for $q = M_Z$. In Table (\ref{tab:KsvnsvmZ}), we compare the K-factor values of SV and SV+NSV resummed predictions at various orders for benchmark rapidity values. We find that there is an increment of 3.15\%,  2.75\% and 0.625\% in the rapidity distribution when going from LL to $\overline{\rm{LL}}$, NLL to $\overline{\rm{NLL}}$ and NNLL to $\overline{\rm{NNLL}}$, respectively, at the central rapidity region. Fig.(\ref{fig:KsvnsvMz}) demonstrates this trend for a wider range of rapidity values. In addition, the K-factors curves of $\overline{\rm{NLL}}$ and $\overline{\rm{NNLL}}$ almost overlap with each other for a wide range of rapidity values which is not observed for the case of NLL and NNLL curves. This suggests that there is better perturbative convergence when the NSV logarithms are taken into account.

We now turn to study the scale uncertainties arising from SV+NSV resmmation in comparison to the SV resummation. We first analyse the behaviour of both SV and SV+NSV resummed rapidity distributions as a function of $y$ under the 7-point scale scale variation as depicted in fig.(\ref{fig:7pt2Tsvnsv}) for $q=2$ TeV. We observe that inclusion of SV as well as SV+NSV resummed corrections reduces the uncertainty of fixed order results at both NLO and NNLO accuracy. This reductions in the uncertainty is prominent for lower rapidity values $|y| \leq 0.5$ as shown in the insets in fig.(\ref{fig:7pt2Tsvnsv}). As can be seen from table (\ref{tab:svnsv2T}), the uncertainty at NLO+NLL is comparable to that of NLO+$\overline{\rm NLL}$ around the central rapidity region. However, the uncertainty at NNLO+NNLL gets worse when we add the resummed NSV contributions at that accuracy. For instance, the uncertainty at NNLO+NNLL lies in the range (-0.34\%,+0.23\%) whereas it is increased to (-0.31\%,+0.53\%) at NNLO+$\overline{\rm NNLL}$ for $y=0$. This hint towards our earlier findings in the previous section that the sensitivity of the SV+NSV resummed results to the unphysical scales increases due to the lack of resummed NSV predictions from off-diagonal $qg$ channel. Next we move on to compare the SV and SV+NSV resummed predictions under the variation of each of these scales separately.  
\begin{table*}
 \renewcommand{\arraystretch}{2.7}
\begin{tabular}{ |P{0.6cm}||P{2.0cm}|P{2.2cm}|P{2.3cm}||P{2.3cm}|P{2.3cm}|P{2.5cm}|}
 \hline
  y &NLO&NLO+NLL&NLO+${\overline{\rm NLL}}$ &NNLO&NNLO+NNLL&NNLO+${\overline{\rm NNLL}}$\\
 \hline
\hline
 0 & 3.323$^{+2.836 \%}_{-2.928\%}$&  3.3927$^{+1.380\%}_{-1.5263\%}$  &  3.460$^{+1.395\%}_{-1.253\%}$  &  3.4405$^{+0.597\%}_{-0.889\%}$ & 3.4503$^{+0.226\%}_{-0.337\%}$ &  3.470$^{+ 0.533\%}_{-0.312\%}$   \\
 
 0.4 &3.105$^{+2.881 \%}_{-2.998\%}$& 3.1803$^{+1.390\%}_{-1.431\%}$   & 3.244$^{+ 1.408\%}_{-1.268\%}$  &  3.2328$^{+0.600\%}_{-0.919\%}$    &  3.2462$^{+ 0.204\%}_{-0.329\%}$  &3.2660 $^{+0.596\%}_{-0.397\%}$  \\
 
 0.8 &2.295$^{+ 3.079\%}_{-3.290\%}$&  2.3755$^{+ 1.427\%}_{-1.281\%}$ & 2.426 $^{+1.563\%}_{-1.322\%}$    &   2.4096$^{+0.647\%}_{-1.033\%}$  & 2.4298$^{+0.308\%}_{-0.276\%}$& 2.446$^{+0.790\%}_{-0.653\%}$  \\
 
 1.2 &0.9384$^{+ 3.655\%}_{-3.914\%}$& 0.9946$^{+1.528\%}_{-1.399\%}$    &  1.0192$^{+2.368\%}_{-1.458\%}$   &  0.9865$^{+0.788\%}_{-1.266\%}$ &1.0044$^{+ 0.725\%}_{-0.583\%}$  &1.0131$^{+1.380\%}_{-1.206\%}$ \\
\hline
\end{tabular}
\caption{Fixed order, SV and SV+NSV resummed cross sections in $10^{-7}$ pb/GeV with 7-point scale uncertainties in \% around the central scale $\mu_R=\mu_F=2$ TeV.} \label{tab:svnsv2T}
\end{table*}

\begin{figure*}
\includegraphics[scale=0.55]{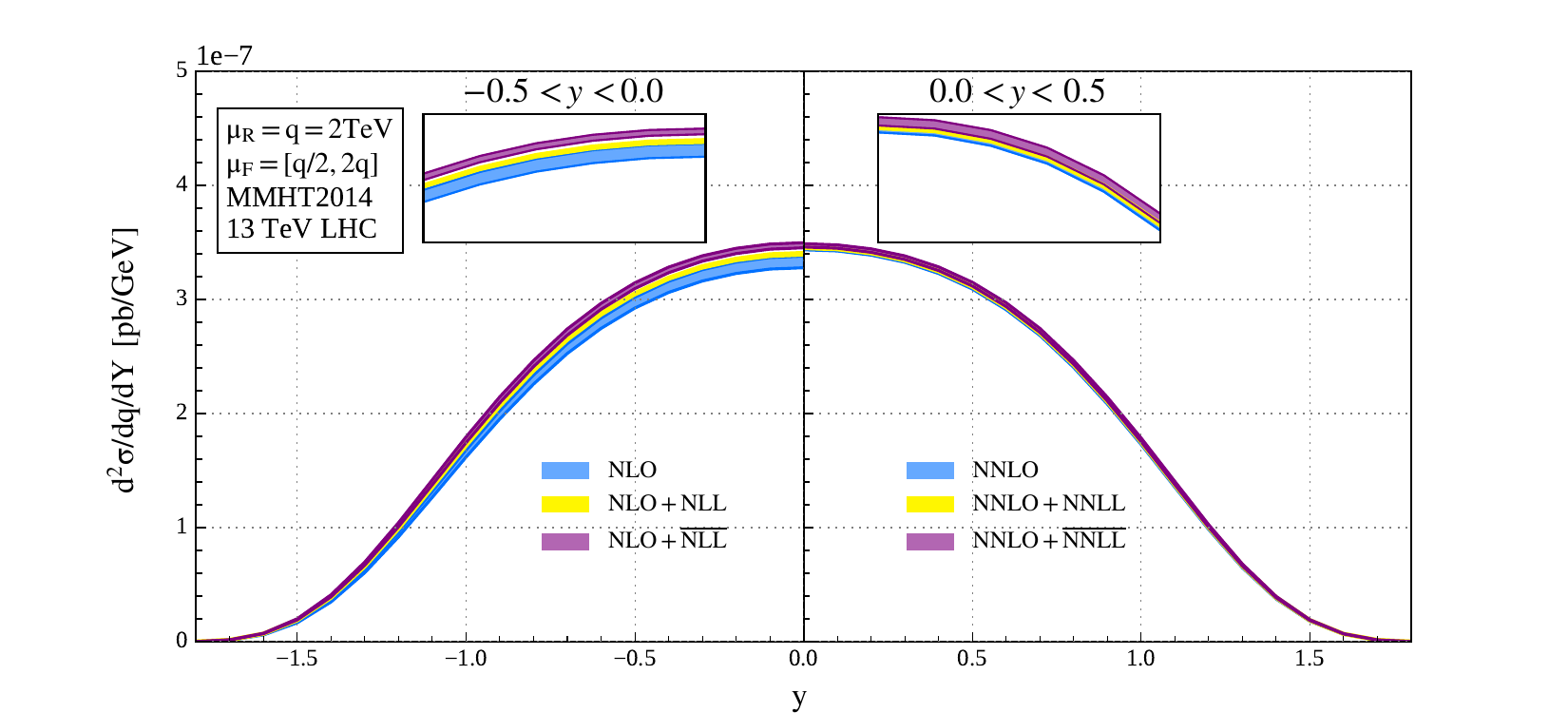}
\caption{Comparison of $\mu_F$ scale variation between SV and SV+NSV resummed results matched to NLO(left panel) and NNLO(right panel) with the scale $\mu_R$ held fixed at $q=2$ TeV.}
\label{fig:muF2Tsvnsv}
\end{figure*}

We first consider the behaviour of both SV and SV+NSV resummed rapidity distributions as a function of $y$ under the $\mu_F$ scale variation with the scale $\mu_R$ fixed at $q=2$ TeV as depicted in fig.(\ref{fig:muF2Tsvnsv}). In general, the bands corresponding to SV+NSV resummed predictions are wider than that of SV predictions over the entire rapidity region. We also find that the width of the bands corresponding to fixed order rapidity distributions gets reduced with the inclusion of both SV (NLL) and SV+NSV ($\overline{\rm NLL}$) resummed corrections at NLO. For instance, the uncertainty is (-1.36\%,+1.7\%) at NLO whereas it is reduced to  (-0\%,0.46\%) and (-0.2\%,+1.07\%) at NLO+NLL and NLO+$\overline{\rm NLL}$ respectively for the central rapidity value. This can be associated with the earlier observation of $q\bar{q}$ and $qg$ contributions at NLO. We have already seen that $q\bar{q}$ is the dominating channel at NLO and hence the uncertainty is expected to get better as we include the resummed corrections coming from that channel. On the other hand, though the uncertainty at NNLO gets improved by the addition of NNLL SV resummed corrections, it gets worse when we include the NSV corrections through $\overline{\rm NNLL}$. These observations can be seen from the insets in fig.(\ref{fig:muF2Tsvnsv}) for lower rapidity values $|y| \leq 0.5$ .   
As we know, the SV resummed terms come only from diagonal $q\bar{q}$ channel, therefore they do not need any compensating factor to reduce its uncertainty. In contrary to this, the NSV resummed predictions which we have included here is incomplete due to missing contributions from off-diagonal $qg$ channel. Consequently the NSV included results will show the residual $\mu_F$ uncertainty due to mixing of various partonic channels. However, the scenario will be different if we keep the scale $\mu_F$ fixed and vary the renormalisation scale $\mu_R$.

 \begin{figure*}
\includegraphics[scale=0.55]{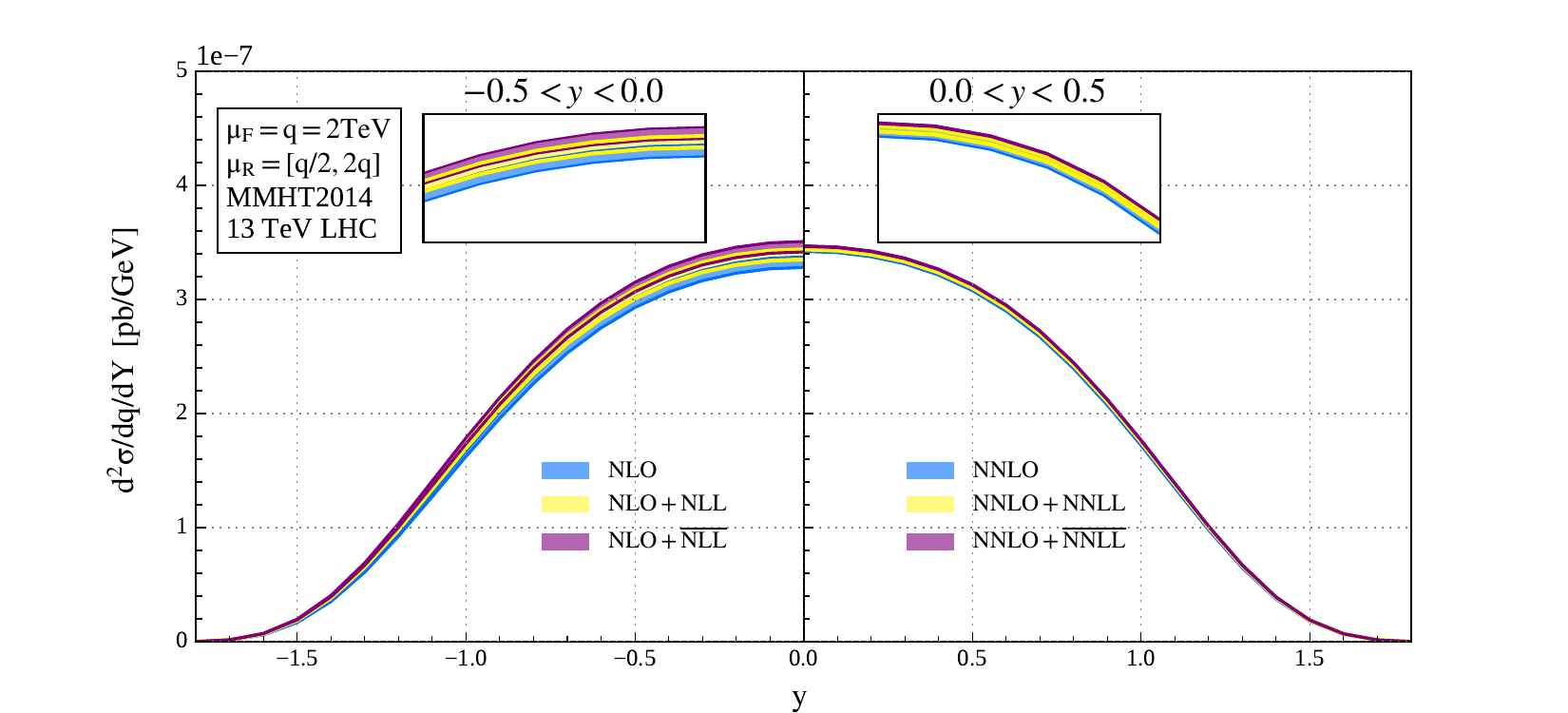}
\caption{Comparison of $\mu_R$ scale variation between SV and SV+NSV resummed results matched to NLO(left panel) and NNLO(right panel) with the scale $\mu_F$ held fixed at $q=2$ TeV.}
\label{fig:muR2Tsvnsv}
\end{figure*}

Fig.(\ref{fig:muR2Tsvnsv}) shows the comparison of SV and SV+NSV resummed results under the $\mu_R$ scale variation for $q=2$ TeV. Here, for the case of $\mu_R$ variation as well, the uncertainties of fixed order results get improved by the inclusion of both SV and NSV resummed corrections.  
Interestingly, at NNLO, the width of the bands gets reduced substantially when the NSV resummed correction at $\overline{\rm NNLL}$ accuracy is added in comparison to its SV counterpart. This improvement by the inclusion of $\overline{\rm NNLL}$ NSV resummed corrections is notable for lower rapidity values $|y| \leq 0.5$ as shown in the insets in fig.(\ref{fig:muR2Tsvnsv}). The uncertainty at NNLO around the central rapidity region lies between (-0.57\%,+0.5\%) which gets reduced to (-0.33\%,+0.22\%) when the SV (NNLL) corrections are added. And, it gets further improved to (-0.16\%,0\%) with the inclusion of NSV corrections ($\overline{\rm NNLL}$). This emphasizes that the resummed NSV contributions play vital role in bringing down the $\mu_R$ scale uncertainty as we go to higher logarithmic corrections.  

In summary, we found that the uncertainty becomes better with the inclusion of both SV (NLL) and NSV ($\overline{\rm NLL}$) resummed corrections at NLO under $\mu_F$ as well as $\mu_R$ scale variations. But at NNLO, under $\mu_F$ scale variation, the inclusion of NSV $\overline{\rm NNLL}$ corrections increases the uncertainty whereas the SV NNLL corrections brings it down significantly. This indicates that the NSV resummed corrections here require the resummed contributions from $qg$ channel as a compensating factor to improve the uncertainty.  Note that in all these analysis, we studied the impact of fixed order and resummed CFs using same PDF sets to desired logarithmic accuracy for both of them.  For studies related to $\mu_F$ variations, it is worthwhile to consider resummed PDFs if they are available. However, as far as the $\mu_R$ uncertainty is concerned, the NSV corrections show nice behaviour especially at NNLO+$\overline{\rm NNLL}$ accuracy with notable reduction in the uncertainty. This suggests that the resummed NSV terms plays substantial role in improving the $\mu_R$ scale uncertainty in comparison to its SV counterpart. 
\section{Discussion and Conclusion}
Through this article, we provide for the first time, the numerical predictions for resummed next-to soft virtual contributions up to NNLO + $\rm \overline{NNLL}$ accuracy to the rapidity distribution of pair of leptons in the Drell-Yan process at the LHC. 
By restricting ourselves to the mechanism where only neutral gauge bosons like photon and $Z$ boson produce leptons, we have used our recent formalism \cite{Ajjath:2020lwb} to systematically resum NSV logarithms to all orders. In our previous work on Drell-Yan inclusive cross-section, we have quantified the significant contribution of the NSV logarithms in the fixed order predictions \cite{Ajjath:2021lvg}. This serves as the motivating factor to study the numerical significance  of these collinear logarithms in the case of rapidity distribution as well. 

We have quantified the numerical effects of SV+NSV higher order predictions by providing the K-factor values for central scale $\mu_R = \mu_F = M_Z$. We find that there is an enhancement of 4.9\%, 3.98\% and 1.24\% at LO+$\rm \overline{LL}$, NLO+$\rm \overline{NLL}$ and NNLO+$\rm \overline{NNLL}$ respectively by the inclusion of SV+NSV resumed results. Also, there is an improvement in the perturbative convergence over the fixed order results till NNLO+$\rm \overline{NNLL}$ accuracy. The sensitivity of our predictions to the unphysical scales $\mu_R$ and $\mu_F$ is studied using the canonical 7-point scale variation approach. We have given the plot of 7-point scale variation for two values of invariant mass, $q=M_Z$ and $q=2$ TeV. We find that at $q=M_Z$, the uncertainty of resummed predictions is more than the corresponding fixed order results till NNLO. However, at $q=2$ TeV, the scale sensitivity at NLO+$\rm \overline{NLL}$ is decreased over the entire rapidity region whereas at NNLO+$\rm \overline{NNLL}$, it gets reduced around the central rapidity region. Thus, by doing a comparative study of the scale uncertainties at two different $q$ values, we infer that the resummation effects become prominent as we go to higher values of $q$. Nevertheless, there is a systematic reduction in the uncertainty of the resummed results while moving to higher logarithmic accuracy for both $q=M_z$ and $q=2$ TeV. 

Further analysis of the scale dependency revealed that the 7-point scale uncertainties of resummed predictions are largely governed by the factorisation scale $\mu_F$ especially at NNLO + $\rm \overline{NNLL}$. Moreover, the comparative study of SV and SV+NSV resummed results shows that the NSV part of the resummation increases the uncertainty due to $\mu_F$ scale variations. We know that different partonic channels mix under factorisation scale variations when they are convoluted with appropriate PDFs. Therefore, the absence of NSV contributions coming from the off-diagonal $qg$ channel increases the sensitivity to $\mu_F$ scale at the hadronic level. However, this missing compensation from $qg$ channel is more evident at NNLO level due to considerable contribution from $qg$ channel at this order.
This suggests that the NSV resummation corresponding to $qg$ channel is necessary to improve the predictions as we go to higher orders in perturbaton theory. In addition, as far as the $\mu_F$ scale variation is concerned, resummed PDFs are also useful to be included for better results.

The independent study of renormalisation scale variation shows that the improvement in the scale uncertainty at NLO + $\rm \overline{NLL}$ is not quantitatively significant, however at NNLO + $\rm \overline{NNLL}$, there is a substantial decrease in $\mu_R$ scale sensitivity as compared to the corresponding fixed order results. Note that the $\mu_R$ scale uncertainty at NNLO is reduced from (-0.56\%, +0.5\%) to (-0.16\%, 0\%) for the central rapidity region by the inclusion of $\overline{\rm NNLL}$.  
From the comparison of SV and SV+NSV resummed results, we find that it is the inclusion of NSV resummed corrections at $\rm \overline{NNLL}$ accuracy to its SV counterpart which brings down the $\mu_R$ scale dependency to a great extent. This is expected because different channels being renormalisation group invariant, do not mix under $\mu_R$ scale variation.  

\section{Acknowledgements}

We thank J. Michel and F. Tackmann for
third order DY results of rapidity for comparing purposes and 
C. Duhr and B. Mistlberger for providing third order results for the inclusive reactions. 
Further, we are grateful to Marco Bonvini for notifying some typos in the paper.
A. A. H is supported by the French ANR under the grant ANR-20-CE31-0015 (“PrecisOnium”). In addition we would also like to thank the computer administrative unit of IMSc  for their help and support.

\appendix
\begin{widetext}
\section{NSV Resummation exponents $   \overline{g}_{d,i}^q(\omega)$} \label{app:gbdN}
The NSV resummation exponents $   \overline{g}_{d,i}^q(\omega)$ given in \eqref{eq:PsiN} are provided below. 
\begin{align}
\begin{autobreak}
   \overline{g}_{d,1}^q(\omega) =

       \frac{1}{\beta_0} C_F   \bigg\{
           2 L_\omega
          \bigg\}\,,

\end{autobreak}\\
  \begin{autobreak}

   \overline{g}_{d,2}^q(\omega) =

       \frac{1}{1- \omega} \Bigg[ \frac{1}{\beta_0^2} C_F C_A n_f   \bigg\{
          - \frac{20}{3} \omega 
          - \frac{20}{3} L_\omega 
          \bigg\}

       +\frac{1}{\beta_0^2} C_F C_A^2   \bigg\{
          \frac{68}{3} \omega
          + \frac{68}{3} L_\omega
          \bigg\}

       +\frac{1}{\beta_0^2} C_F^2 n_f   \bigg\{
          - 4 \omega 
          - 4 L_\omega 
          \bigg\}

       +\frac{1}{\beta_0} C_F n_f  \bigg\{
           \frac{20}{9} \omega 
          \bigg\}

       +\frac{1}{\beta_0} C_F C_A   \bigg\{
          - \frac{134}{9} \omega
          + 4 \omega \zeta_2
          \bigg\}

       + C_F   \bigg\{
          - 2
          + 2 L_{qr}
          - 2 L_{fr}
          + 2 L_{fr} \omega
          - 4 \gamma_E
          \bigg\} \Bigg] \,,

\end{autobreak}\\
  \begin{autobreak}

   \overline{g}_{d,3}^q(\omega) =

       \frac{1}{(1- \omega)^2} \Bigg[ \frac{1}{\beta_0^3} C_F C_A^2 n_f^2  \bigg\{
           \frac{100}{9} \omega^2 
          - \frac{100}{9} L_\omega^2 
          \bigg\}

       +\frac{1}{\beta_0^3} C_F C_A^3 n_f   \bigg\{
          - \frac{680}{9} \omega^2 
          + \frac{680}{9} L_\omega^2 
          \bigg\}

       +\frac{1}{\beta_0^3} C_F C_A^4   \bigg\{
          \frac{1156}{9} \omega^2
          - \frac{1156}{9} L_\omega^2
          \bigg\}

       +\frac{1}{\beta_0^3} C_F^2 C_A  n_f^2  \bigg\{
          \frac{40}{3} \omega^2 
          - \frac{40}{3} L_\omega^2 
          \bigg\}

       +\frac{1}{\beta_0^3} C_F^2 C_A^2 n_f   \bigg\{
          - \frac{136}{3} \omega^2 
          + \frac{136}{3} L_\omega^2 
          \bigg\}

       +\frac{1}{\beta_0^3} C_F^3 n_f^2   \bigg\{
           4 \omega^2 
          - 4 L_\omega^2 
          \bigg\}

       +\frac{1}{\beta_0^2} C_F C_A n_f^2  \bigg\{
           \frac{200}{27} \omega 
          - \frac{31}{6} \omega^2 
          + \frac{200}{27} L_\omega 
          \bigg\}

       +\frac{1}{\beta_0^2} C_F C_A^2 n_f  \bigg\{
          - \frac{2020}{27} \omega 
          + \frac{40}{3} \omega \zeta_2 
          + \frac{1145}{18} \omega^2 
          - \frac{20}{3} \omega^2 \zeta_2 
          - \frac{2020}{27} L_\omega 
          + \frac{40}{3} L_\omega \zeta_2 
          \bigg\}

       +\frac{1}{\beta_0^2} C_F C_A^3   \bigg\{
           \frac{4556}{27} \omega
          - \frac{136}{3} \omega \zeta_2
          - \frac{2471}{18} \omega^2
          + \frac{68}{3} \omega^2 \zeta_2
          + \frac{4556}{27} L_\omega
          - \frac{136}{3} L_\omega \zeta_2
          \bigg\}

       +\frac{1}{\beta_0^2} C_F^2 n_f^2  \bigg\{
           \frac{40}{9} \omega 
          - \frac{31}{9} \omega^2 
          + \frac{40}{9} L_\omega 
          \bigg\}

       +\frac{1}{\beta_0^2} C_F^2 C_A n_f  \bigg\{
          - \frac{268}{9} \omega
          + 8 \omega \zeta_2 
          + \frac{473}{18} \omega^2 
          - 4 \omega^2 \zeta_2 
          - \frac{268}{9} L_\omega 
          + 8 L_\omega \zeta_2 
          \bigg\}

       +\frac{1}{\beta_0^2} C_F^3 n_f  \bigg\{
          - \omega^2 
          \bigg\}

       +\frac{1}{\beta_0} C_F n_f^2  \bigg\{
           \frac{8}{27} \omega 
          - \frac{4}{27} \omega^2 
          \bigg\}

       +\frac{1}{\beta_0} C_F C_A n_f   \bigg\{
           \frac{418}{27} \omega 
          + \frac{56}{3} \omega  \zeta_3
          - \frac{80}{9} \omega \zeta_2 
          - \frac{209}{27} \omega^2 
          - \frac{28}{3} \omega^2  \zeta_3
          + \frac{40}{9} \omega^2 \zeta_2 
          - \frac{20}{3} L_\omega 
          + \frac{20}{3} L_\omega L_{qr} 
          - \frac{40}{3} L_\omega \gamma_E 
          \bigg\}

       +\frac{1}{\beta_0} C_F C_A^2   \bigg\{
          - \frac{245}{3} \omega
          - \frac{44}{3} \omega \zeta_3
          + \frac{536}{9} \omega \zeta_2
          - \frac{88}{5} \omega \zeta_2^2
          + \frac{245}{6} \omega^2
          + \frac{22}{3} \omega^2 \zeta_3
          - \frac{268}{9} \omega^2 \zeta_2
          + \frac{44}{5} \omega^2 \zeta_2^2
          + \frac{68}{3} L_\omega
          - \frac{68}{3} L_\omega L_{qr}
          + \frac{136}{3} L_\omega \gamma_E
          \bigg\}

       +\frac{1}{\beta_0} C_F^2 n_f  \bigg\{
           \frac{55}{3} \omega 
          - 16 \omega  \zeta_3
          - \frac{55}{6} \omega^2 
          + 8 \omega^2  \zeta_3
          - 4 L_\omega 
          + 4 L_\omega L_{qr} 
          - 8 L_\omega \gamma_E 
          \bigg\}

       + C_F n_f  \bigg\{
          \frac{116}{27} 
          - \frac{32}{9} L_{qr}
          + \frac{2}{3} L_{qr}^2 
          + \frac{20}{9} L_{fr} 
          - \frac{40}{9} L_{fr} \omega 
          + \frac{20}{9} L_{fr} \omega^2 
          - \frac{2}{3} L_{fr}^2 
          + \frac{4}{3} L_{fr}^2 \omega 
          - \frac{2}{3} L_{fr}^2 \omega^2 
          + \frac{64}{9} \gamma_E 
          - \frac{8}{3} \gamma_E L_{qr} 
          + \frac{8}{3} \gamma_E^2 
          \bigg\}

       + C_F C_A   \bigg\{
          - \frac{806}{27}
          + 14 \zeta_3
          + 4 \zeta_2
          + \frac{200}{9} L_{qr}
          - 4 L_{qr} \zeta_2
          - \frac{11}{3} L_{qr}^2
          - \frac{134}{9} L_{fr}
          + 4 L_{fr} \zeta_2
          + \frac{268}{9} L_{fr} \omega
          - 8 L_{fr} \omega \zeta_2
          - \frac{134}{9} L_{fr} \omega^2
          + 4 L_{fr} \omega^2 \zeta_2
          + \frac{11}{3} L_{fr}^2
          - \frac{22}{3} L_{fr}^2 \omega
          + \frac{11}{3} L_{fr}^2 \omega^2
          - \frac{400}{9} \gamma_E
          + 8 \gamma_E \zeta_2
          + \frac{44}{3} \gamma_E L_{qr}
          - \frac{44}{3} \gamma_E^2
          \bigg\}\Bigg]\,.
\end{autobreak}
\end{align}

\section{NSV Resummation exponents $  h^q_{d,ij}(\omega)$ and $\tilde h^q_{d,ii}(\omega,\omega_l)$} \label{app:hdN}
The NSV resummation exponents $h^q_{d,ij}(\omega)$ and $\tilde h^q_{d,ii}(\omega,\omega_l)$ given in \eqref{hg} are provided below. 
\begin{align}
\begin{autobreak}  
   h^q_{d,00}(\omega) =

       \frac{1}{\beta_0} C_F   \bigg\{
          - 4 L_\omega
          \bigg\}\,, 
\qquad
   h^q_{d,01}(\omega) = 0\,,
\end{autobreak}\\ 
  \begin{autobreak}
   h^q_{d,10}(\omega) =

       \frac{1}{2 \beta_0^2 (\omega -1)} \Bigg[ \beta_1 C_F   \bigg\{
           8 \omega
          + 8 L_\omega
          \bigg\}

       + \beta_0 C_F n_f   \bigg\{
           \frac{80}{9} \omega
          \bigg\}

       + \beta_0 C_F C_A   \bigg\{
          - \frac{536}{9} \omega
          + 16 \omega \zeta_2
          \bigg\}

       + \beta_0 C_F^2   \bigg\{
           24 \omega
          - 32 \gamma_E \omega
          \bigg\}

       + \beta_0^2 C_F   \bigg\{
          - 4
          - 8 \omega
          - 8 L_{fr}
          + 8 L_{fr} \omega
          + 8 L_{qr}
          - 16 \gamma_E
          \bigg\} \Bigg]\,,
   
\end{autobreak}\\ 
  \begin{autobreak}
  
  \tilde{h}^q_{d,11}(\omega, \omega_l) = 
  
            \frac{ C_F^2}{\beta_0 }   \bigg\{
           -\frac{4 \omega_l}{(\omega -1)^2}
          - \frac{16 \omega}{(\omega -1)} 
          \bigg\}\,,
   \end{autobreak}\\ 
  \begin{autobreak}
   h^q_{d,20}(\omega) =

       \frac{1}{2 \beta_0^3 (\omega -1)^2} \Bigg[ \beta_1^2 C_F   \bigg\{
          - 4 \omega^2
          + 4 L_\omega^2
          \bigg\}

       + \beta_0 \beta_2 C_F   \bigg\{
           4 \omega^2
          \bigg\}

       + \beta_0 \beta_1 C_F n_f   \bigg\{
           \frac{80}{9} \omega
          - \frac{40}{9} \omega^2
          + \frac{80}{9} L_\omega
          \bigg\}

       + \beta_0 \beta_1 C_F C_A   \bigg\{
          - \frac{536}{9} \omega
          + 16 \omega \zeta_2
          + \frac{268}{9} \omega^2
          - 8 \omega^2 \zeta_2
          - \frac{536}{9} L_\omega
          + 16 L_\omega \zeta_2
          \bigg\}

       + \beta_0 \beta_1 C_F^2   \bigg\{
           24 \omega
          - 12 \omega^2
          - 32 \gamma_E \omega
          + 16 \gamma_E \omega^2
          + 24 L_\omega
          - 32 L_\omega \gamma_E
          \bigg\}

       + \beta_0^2 C_F n_f^2   \bigg\{
          - \frac{32}{27} \omega
          + \frac{16}{27} \omega^2
          \bigg\}

       + \beta_0^2 C_F C_A n_f   \bigg\{
          - \frac{1672}{27} \omega
          - \frac{224}{3} \omega \zeta_3
          + \frac{320}{9} \omega \zeta_2
          + \frac{836}{27} \omega^2
          + \frac{112}{3} \omega^2 \zeta_3
          - \frac{160}{9} \omega^2 \zeta_2
          \bigg\}

       + \beta_0^2 C_F C_A^2   \bigg\{
           \frac{980}{3} \omega
          + \frac{176}{3} \omega \zeta_3
          - \frac{2144}{9} \omega \zeta_2
          + \frac{352}{5} \omega \zeta_2^2
          - \frac{490}{3} \omega^2
          - \frac{88}{3} \omega^2 \zeta_3
          + \frac{1072}{9} \omega^2 \zeta_2
          - \frac{176}{5} \omega^2 \zeta_2^2
          \bigg\}

       + \beta_0^2 C_F^2 n_f   \bigg\{
          - 44 \omega
          + 64 \omega \zeta_3
          + \frac{64}{3} \omega \zeta_2
          + 22 \omega^2
          - 32 \omega^2 \zeta_3
          - \frac{32}{3} \omega^2 \zeta_2
          - \frac{640}{9} \gamma_E \omega
          + \frac{320}{9} \gamma_E \omega^2
          \bigg\}

       + \beta_0^2 C_F^2 C_A   \bigg\{
          - \frac{604}{3} \omega
          + 96 \omega \zeta_3
          - \frac{208}{3} \omega \zeta_2
          + \frac{302}{3} \omega^2
          - 48 \omega^2 \zeta_3
          + \frac{104}{3} \omega^2 \zeta_2
          + \frac{4288}{9} \gamma_E \omega
          - 128 \gamma_E \omega \zeta_2
          - \frac{2144}{9} \gamma_E \omega^2
          + 64 \gamma_E \omega^2 \zeta_2
          \bigg\}

       + \beta_0^2 C_F^3   \bigg\{
          - 12 \omega
          - 192 \omega \zeta_3
          + 96 \omega \zeta_2
          + 6 \omega^2
          + 96 \omega^2 \zeta_3
          - 48 \omega^2 \zeta_2
          \bigg\}

       + \beta_0^2 \beta_1 C_F   \bigg\{
          - 12 L_\omega
          + 8 L_\omega L_{qr}
          - 16 L_\omega \gamma_E
          \bigg\}

       + \beta_0^3 C_F n_f   \bigg\{
          - \frac{536}{27}
          + \frac{32}{3} \zeta_2
          - \frac{80}{9} \omega
          + \frac{40}{9} \omega^2
          - \frac{80}{9} L_{fr}
          + \frac{160}{9} L_{fr} \omega
          - \frac{80}{9} L_{fr} \omega^2
          + \frac{80}{9} L_{qr}
          - \frac{160}{9} \gamma_E
          \bigg\}

       + \beta_0^3 C_F C_A   \bigg\{
           \frac{2000}{27}
          - 56 \zeta_3
          - \frac{200}{3} \zeta_2
          + \frac{536}{9} \omega
          - 16 \omega \zeta_2
          - \frac{268}{9} \omega^2
          + 8 \omega^2 \zeta_2
          + \frac{536}{9} L_{fr}
          - 16 L_{fr} \zeta_2
          - \frac{1072}{9} L_{fr} \omega
          + 32 L_{fr} \omega \zeta_2
          + \frac{536}{9} L_{fr} \omega^2
          - 16 L_{fr} \omega^2 \zeta_2
          - \frac{536}{9} L_{qr}
          + 16 L_{qr} \zeta_2
          + \frac{892}{9} \gamma_E
          - 32 \gamma_E \zeta_2
          \bigg\}

       + \beta_0^3 C_F^2   \bigg\{
          - 8 \zeta_2
          - 24 L_{fr}
          + 48 L_{fr} \omega
          - 24 L_{fr} \omega^2
          + 24 L_{qr}
          - 28 \gamma_E
          + 32 \gamma_E L_{fr}
          - 64 \gamma_E L_{fr} \omega
          + 32 \gamma_E L_{fr} \omega^2
          - 32 \gamma_E L_{qr}
          + 56 \gamma_E^2
          \bigg\}

       + \beta_0^4 C_F   \bigg\{
           16 \zeta_2
          + 8 L_{fr}
          - 16 L_{fr} \omega
          + 8 L_{fr} \omega^2
          - 4 L_{fr}^2
          + 8 L_{fr}^2 \omega
          - 4 L_{fr}^2 \omega^2
          - 12 L_{qr}
          + 4 L_{qr}^2
          + 24 \gamma_E
          - 16 \gamma_E L_{qr}
          + 16 \gamma_E^2
          \bigg\} \Bigg] \,,
   
\end{autobreak}\\ 
  \begin{autobreak}
   h^q_{d,21}(\omega) =

        \frac{1}{2 \beta_0^2 (\omega -1)^2}  \Bigg[ \beta_1 C_F^2   \bigg\{
          - 32 \omega
          + 16 \omega^2
          - 32 L_\omega
          \bigg\}

       + \beta_0 C_F^2 n_f   \bigg\{
          - \frac{640}{9} \omega
          + \frac{320}{9} \omega^2
          \bigg\}

       + \beta_0 C_F^2 C_A   \bigg\{
           \frac{4288}{9} \omega
          - 128 \omega \zeta_2
          - \frac{2144}{9} \omega^2
          + 64 \omega^2 \zeta_2
          \bigg\}

       + \beta_0^2 C_F C_A   \bigg\{
          - 20
          \bigg\}

       + \beta_0^2 C_F^2   \bigg\{
           20
          + 32 L_{fr}
          - 64 L_{fr} \omega
          + 32 L_{fr} \omega^2
          - 32 L_{qr}
          + 48 \gamma_E
          \bigg\} \Bigg]\,,
          \end{autobreak} 
          \\
 \begin{autobreak}
 \tilde{h}^q_{d,22}(\omega, \omega_l) = 
         \frac{\omega_l}{\beta_0 (\omega -1)^3} \Bigg[ C_F^2 n_f   \bigg\{
          \frac{32}{27} 
          \bigg\} + 
        C_F^2 C_A   \bigg\{
          - \frac{176}{27}
          \bigg\} \Bigg] \,,
\end{autobreak}
 \end{align}

where $\gamma_E$ is the Euler-Mascheroni constant. Here, ${L}_{\omega}=\ln(1-\omega)$ with $\omega = \beta_0 a_s(\mu_R^2) \ln N_1 N_2$, $\omega_l = \beta_0 a_s(\mu_R^2) \ln N_l$ with $l=1,2$, $L_{qr} = \ln \big(\frac{q^2}{\mu_R^2}\big)$ and $L_{fr} = \ln \big(\frac{\mu_F^2}{\mu_R^2}\big)$.
\end{widetext}

\bibliography{dyres}
\bibliographystyle{apsrev4-1}
\end{document}